\documentclass[twocolumn]{article}
\usepackage[final,authormarkup=none,commentmarkup=footnote,deletedmarkup=colored]{changes}
\setdeletedmarkup{\textcolor{black!50}{#1}}
\usepackage[super]{nth}
\usepackage{graphicx}
\usepackage{amsmath}
\usepackage[left=1.5cm,right=1.5cm,top=2cm,bottom=2cm]{geometry}

\usepackage[para]{threeparttable}
\usepackage{xfrac}
\usepackage{interval}
\usepackage[export]{adjustbox}
\usepackage{xfp}
\usepackage{amssymb}
\usepackage{multirow}
\usepackage{microtype}
\usepackage{lineno}
\usepackage{mathtools}
\usepackage{stmaryrd}
\usepackage{colortbl}
\usepackage{mathdots}
\usepackage{siunitx}
\usepackage{booktabs}
\usepackage{pgfplots}
\usepackage[doi=false,eprint=false,isbn=false,maxnames=6,style=ieee]{biblatex}
\pgfplotsset{compat=1.17}
\pgfplotsset{colormap={inferno}{%
rgb = (1.46200e-03, 4.66000e-04, 1.38660e-02)
rgb = (2.94320e-02, 2.15030e-02, 1.14621e-01)
rgb = (9.29900e-02, 4.55830e-02, 2.34358e-01)
rgb = (1.83429e-01, 4.03290e-02, 3.54971e-01)
rgb = (2.71347e-01, 4.09220e-02, 4.11976e-01)
rgb = (3.60284e-01, 6.92470e-02, 4.31497e-01)
rgb = (4.41207e-01, 9.93380e-02, 4.31594e-01)
rgb = (5.28444e-01, 1.30341e-01, 4.18142e-01)
rgb = (6.09330e-01, 1.59474e-01, 3.93589e-01)
rgb = (6.94627e-01, 1.95021e-01, 3.54388e-01)
rgb = (7.69556e-01, 2.36077e-01, 3.07485e-01)
rgb = (8.41969e-01, 2.92933e-01, 2.48564e-01)
rgb = (8.98192e-01, 3.58911e-01, 1.88860e-01)
rgb = (9.44285e-01, 4.42772e-01, 1.20354e-01)
rgb = (9.72590e-01, 5.29798e-01, 5.33240e-02)
rgb = (9.86964e-01, 6.30485e-01, 3.09080e-02)
rgb = (9.84865e-01, 7.28427e-01, 1.20785e-01)
rgb = (9.66243e-01, 8.36191e-01, 2.61534e-01)
rgb = (9.46392e-01, 9.30761e-01, 4.42367e-01)
rgb = (9.88362e-01, 9.98364e-01, 6.44924e-01)
}}
\newcommand{\drawcolorbar}{%
	\pgfplotscolorbardrawstandalone[
		scale=0.32, colormap={example}{samples of colormap = (8 of inferno)},
		colorbar horizontal,point meta max=0.2,colorbar style={ticks=none},
	]%
}
\usepackage[ruled,linesnumbered,noend]{algorithm2e}
\usepackage{tikz}
\usepackage{glossaries}
\usepackage{ifthen}
\usetikzlibrary{shapes.arrows,shapes.geometric,spy,backgrounds,calc,matrix}
\newacronym{ai}{{AI}}{Artificial Intelligence}
\newacronym{art}{{ART}}{Algebraic Reconstruction Technique}
\newacronym{sirt}{{SIRT}}{Simultaneous Iterative Reconstruction Technique}
\newacronym{sart}{{SART}}{Simultaneous Algebraic Reconstruction Technique}
\newacronym{bicav}{{BICAV}}{Block Iterative Component Averaging}
\newacronym{cgls}{{CGLS}}{Conjugate Gradient Least Squares}
\newacronym{ossqs}{{OS-SQS}}{Ordered Subset Separable Quadratic Surrogates}
\newacronym{ann}{{ANN}}{Artificial Neural Network}
\newacronym{bp}{{BP}}{Backprojection}
\newacronym{admm}{{ADMM}}{Alternating Direction Method of Multipliers}
\newacronym{cnn}{{CNN}}{convolutional neural network}
\newacronym{cg}{{CG}}{Conjugate Gradient}
\newacronym{cs}{{CS}}{compressed sensing}
\newacronym{ct}{{CT}}{computed tomography}
\newacronym{cpu}{{CPU}}{Central Processing Unit}
\newacronym{dicom}{{DICOM}}{Digital Imaging and Communications in Medicine}
\newacronym{dof}{{DOF}}{Degrees of Freedom}
\newacronym{ddr}{{DDR}}{Digitally Reconstructed Radiograph}
\newacronym{dsc}{{DSC}}{Dice Similarity Coefficient}
\newacronym{dti}{{DTI}}{Diffusion Tensor Imaging}
\newacronym{ecg}{{ECG}}{Electrocardiography}
\newacronym{em}{{EM}}{Expectation Maximization}
\newacronym{ft}{{FT}}{Fourier Transform}
\newacronym{fft}{{FFT}}{fast Fourier transform}
\newacronym{fista}{{FISTA}}{Fast Iterative Shrinkage and Thresholding Algorithm}
\newacronym{fbp}{{FBP}}{Filtered Back-Projection}
\newacronym{foe}{{FoE}}{Fields of Experts}
\newacronym{fov}{{FoV}}{Field of View}
\newacronym{gac}{{GAC}}{Geodesic Active Contours}
\newacronym{gan}{{GAN}}{generative adversarial network}
\newacronym{gd}{{GD}}{Gradient Descent}
\newacronym{gmm}{{GMM}}{Gaussian Mixture Model}
\newacronym{gpu}{{GPU}}{Graphics Processing Unit}
\newacronym{hu}{{HU}}{Hounsfield Units}
\newacronym{ista}{{ISTA}}{Iterative Shrinkage and Thresholding Algorithm}
\newacronym{iipg}{{IIPG}}{Inertial Incremental Proximal Gradient}
\newacronym{ipalm}{{iPALM}}{inertial proximal alternating linearized minimization}
\newacronym{lsc}{{l.s.c.}}{lower-semicontinuous}
\newacronym{lista}{{LISTA}}{Learned Iterative Shrinkage and Thresholding Algorithm}
\newacronym{lbfgs}{{L-BFGS}}{Limited-Memory Broyden-Fletcher-Goldfarb-Shanno}
\newacronym{map}{{MAP}}{maximum a-posteriori}
\newacronym{mlp}{{MLP}}{Multi Layer Perceptron}
\newacronym{mr}{{MR}}{Magnetic Resonance}
\newacronym{ml}{{ML}}{Maximum Likelihood}
\newacronym{mri}{{MRI}}{magnetic resonance imaging}
\newacronym{mae}{{MAE}}{Mean Absolute Error}
\newacronym{mse}{{MSE}}{Mean Squared Error}
\newacronym{msssim}{{MS-SSIM}}{Multi-Scale Structural Similarity Index}
\newacronym{ncc}{{NCC}}{Normalized Cross Correlation}
\newacronym{nlm}{{NLM}}{Non-Local Means}
\newacronym{nufft}{{NUFFT}}{non-uniform fast Fourier transform}
\newacronym{nrmse}{{NRMSE}}{Normalized Root Mean Squared Error}
\newacronym{icp}{{ICP}}{Iterative Closest Point}
\newacronym{pat}{{PAT}}{Photoacoustic Tomography}
\newacronym{pca}{{PCA}}{Principal Component Analysis}
\newacronym{pet}{{PET}}{Positron Emission Tomography}
\newacronym{psf}{{PSF}}{Point Spread Function}
\newacronym{psnr}{{PSNR}}{peak signal-to-noise ratio}
\newacronym{rbf}{{RBF}}{Gaussian radial basis function}
\newacronym{relu}{{ReLU}}{Rectified Linear Unit}
\newacronym{roi}{{ROI}}{Region Of Interest}
\newacronym{snr}{{SNR}}{Signal-to-Noise Ratio}
\newacronym{sota}{SotA}{state-of-the-art}
\newacronym{spect}{{SPECT}}{Single Photon Emission Computed Tomography}
\newacronym{ssim}{{SSIM}}{structural similarity}
\newacronym{tof}{{ToF}}{Time of Flight}
\newacronym{tgv}{{TGV}}{Total Generalized Variation}
\newacronym{tv}{{TV}}{total variation}
\newacronym{us}{{US}}{Ultrasound}
\newacronym{vn}{{VN}}{variational network}
\newacronym{sbp}{{SBP}}{Simple Back-Projection}
\newacronym{fdk}{{FDK}}{Feldkamp-Davis-Kress}
\newacronym{aec}{{AEC}}{Automatic Exposure Control}
\newacronym{bm3d}{{BM3D}}{Block Matching and 3D Filtering}
\newacronym{mcmc}{{MCMC}}{Markov chain Monte Carlo}
\newacronym{cd}{{CD}}{Contrastive Divergence}
\newacronym{poe}{{PoE}}{Products of Experts}
\newacronym{mala}{{MALA}}{Metropolis adjusted Langevin algorithm}
\newacronym{lmc}{{LMC}}{Langevin Monte Carlo}
\newacronym{ula}{{ULA}}{unadjusted Langevin algorithm}
\newacronym{tdv}{{TDV}}{Total Deep Variation}
\newacronym{ebm}{{EBM}}{energy-based model}
\newacronym{rss}{{RSS}}{root-sum-of-squares}
\newacronym{sde}{{SDE}}{stochastic differential equation}
\newacronym{nmse}{{NMSE}}{normalized mean-squared error}
\newacronym{acl}{{ACL}}{auto-calibration lines}
\newacronym{mmse}{{MMSE}}{minimum mean-squared-error}
\newacronym{sense}{{SENSE}}{sensitivity encoding}
\newacronym{pi}{{PI}}{parallel imaging}
\newacronym{zf}{{ZF}}{zero-filled}
\newacronym{corpd}{{CORPD}}{coronal proton-density weighted}
\newacronym{corpdfs}{{CORPD-FS}}{coronal proton-density weighted fat-suppressed}

\DeclareMathOperator*{\argmin}{arg\,min}

\DeclareMathOperator*{\prox}{prox}
\newcommand{\R}{\mathbb{R}}
\newcommand{\Function}[3]{#1\colon#2\to#3}
\newcommand{\set}[1]{\mathcal{#1}}
\newcommand{\optimal}[1]{#1^{*}}
\newcommand{\norm}[1]{\left\lVert#1\right\rVert}
\newcommand{\crop}{\tikz[thick, scale=0.35]{%
		\draw [shorten <= -0.4](0.25, 0.25) -- ++(-0.7, 0.);%
		\fill[white] (-0.15, 0.15) rectangle (-0.35, 0.35);%
		\draw (-0.25, -0.25) -- ++(0., 0.7);%

		\draw [shorten <= -0.4] (-0.25, -0.25) -- ++(0.7, 0.);%
		\fill[overlay, white] (0.15, -0.15) rectangle (0.35, -0.35);%
		\draw [shorten >= -2](0.25, 0.25) -- ++(0., -0.5);%
}}
\newcommand{\lrelu}{\tikz[thick, scale=0.35]{%
		\draw (0.0, 0.0) -- ++(0.3, 0.) -- ++(0.5, 0.5);%
}}
\definecolor{elsevierblue}{RGB}{23,179,234}
\usepackage{csquotes}
\usepackage{hyperref}
\usepackage[capitalize]{cleveref}

\title{Stable Deep MRI Reconstruction\\using Generative Priors}
\author{%
	Martin Zach, %
	Florian Knoll, and %
	Thomas Pock%
	\thanks{
		M. Zach and T. Pock are with the Institute of Computer Graphics and Vision, Graz University of Technology, 8010 Graz, Austria (e-mail: \{martin.zach, pock\}@icg.tugraz.at).
	}%
	\thanks{%
		F. Knoll is with with the Department Artificial Intelligence in Biomedical Engineering, Friedrich Alexander University Erlangen Nuernberg. 91052 Erlangen, Germany (e-mail: florian.knoll@fau.de).
	}
}
\begin{document}
\maketitle
\begin{abstract}
	Data-driven approaches recently achieved remarkable success in \gls{mri} reconstruction, but integration into clinical routine remains challenging due to a lack of generalizability and interpretability.
	In this paper, we address these challenges in a unified framework based on generative image priors.
	We propose a novel deep neural network based regularizer which is trained in \replaced{a generative }{an unsupervised} setting on reference magnitude images only.
	After training, the regularizer encodes higher-level domain statistics which we demonstrate by synthesizing images without data.
	Embedding the trained model in a classical variational approach yields high-quality reconstructions irrespective of the sub-sampling pattern.
	In addition, the model shows stable behavior \replaced{when confronted with out-of-distribution data in the form of contrast variation.}{even if the test data deviate significantly from the training data.}
	Furthermore, a probabilistic interpretation provides a distribution of reconstructions and hence allows uncertainty quantification.
	To reconstruct parallel \gls{mri}, we propose a fast algorithm to jointly estimate the image and the sensitivity maps.
	The results demonstrate competitive performance, on par with state-of-the-art end-to-end deep learning methods, while preserving the flexibility with respect to sub-sampling patterns and allowing for uncertainty quantification.
\end{abstract}

\glsresetall{}
\section{Introduction}
A vast literature is dedicated to reducing examination time in \gls{mri} while retaining the diagnostic value of the resulting images.
On the hardware side, \gls{pi}~\cite{pruessmann} exploits spatially varying sensitivity maps of coil arrays.
Such \gls{pi} hardware has become standard in clinical systems, but noise amplification limits the potential speed up with classical reconstruction techniques~\cite{Robson2008}.
On the algorithmic side, \gls{cs} theory~\cite{donoho_compressed_2006} and variational approaches enable greater acceleration under the assumption that the reconstruction has a sparse representation in some basis.
Often, this basis is hand-crafted and only reflects rudimentary aspects of the prior information in the underlying distribution.
As a prominent example, the \gls{tv} assumes sparsity in the spatial image gradients (which translates to piecewise constant image intensities), and has been successfully applied to parallel \gls{mri}~\cite{Knoll2011}.

Hand-crafted prior information, in general, fails to capture the complexity of the underlying distribution~\cite{Huang1999StatisticsON}.
On the other hand, data-driven approaches have shown promising results \gls{mri} reconstruction, often leading to superior results compared to classical \gls{cs} techniques~\cite{hammernik_learning_2017,akcakaya_raki_2019,zbontar_fastmri_2018,putzky_irim_2019,zhou2020dudornet,Narnhofer2019,chung_scoremri_2022}.
Here, information is encoded in a deep neural network, which is trained on reference data in an off-line step.
Data-driven approaches have successfully been applied as pre-processing steps in \( k \)-space~\cite{akcakaya_raki_2019} as well as post-processing steps in image-space~\cite{zbontar_fastmri_2018}.
\Glspl{vn}~\cite{chen_tnrd_2017,hammernik_learning_2017,Kobler2017,cheng_pdnetworks_2019} imitate the structure of an iterative reconstruction scheme by unrolling an optimization algorithm.
In a dynamical setting, this approach has been proven useful for quantifying flow in four-dimensional \gls{mri}~\cite{Vishnevskiy2020}.
At the pinnacle, purely data-driven methods disregard any physical measurement model and instead learn direct maps from \( k \)-space to image-space (AUTOMAP,~\cite{zhu_image_2018}).

\begin{figure*}
	\centering
	\resizebox{\textwidth}{!}{%
	\begin{tikzpicture}
		\def\kspacewidth{2cm}
		\def\kspacexoff{-1.2cm}
		\def\kspaceyoff{0.5cm}
		\def\kspacedx{0.25cm}
		\def\kspacedy{0.25cm}
		\def\sensewidth{0.7cm}
		\def\sensedx{0.1cm}
		\def\sensedy{0.1cm}
		\def\ppad{0.7cm}
		\def\prefix{./figures/examples-coils/01}
		\node (kspace1) at (0 * \kspacedx + \kspacexoff, -1 * \kspacedy + \kspaceyoff) {\includegraphics[width=\kspacewidth]{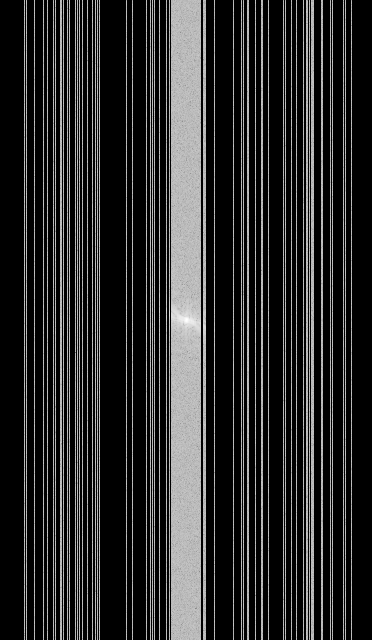}};
		\node at (1 * \kspacedx + \kspacexoff, -2 * \kspacedy + \kspaceyoff) {\includegraphics[width=\kspacewidth]{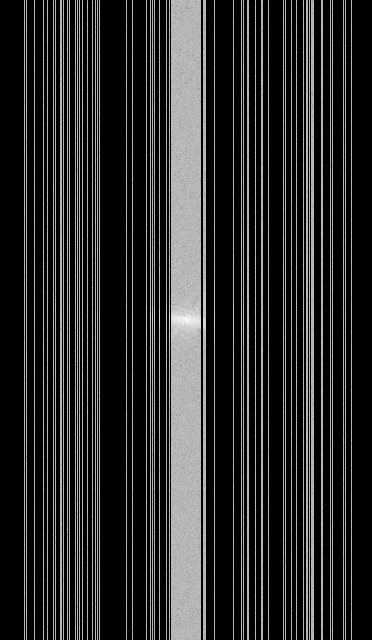}};
		\node [inner sep=0, outer sep=0] (kspace3) at (2 * \kspacedx + \kspacexoff, -3 * \kspacedy + \kspaceyoff) {\includegraphics[width=\kspacewidth]{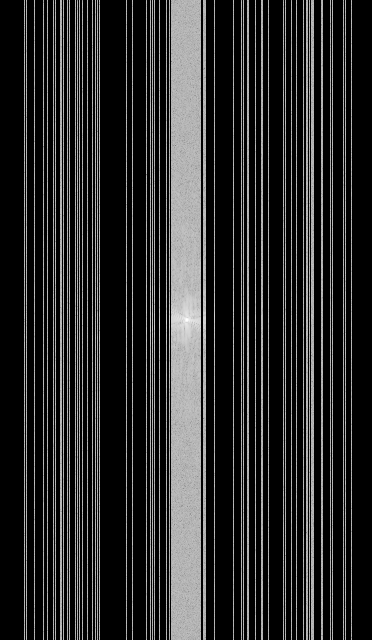}};
		\node at ($(kspace1.north) + (0.2, 0.1)$) {\( z \in \mathbb{C}^m \)};
		\node [inner sep=0, outer sep=0] (kspace6)at (5 * \kspacedx + \kspacexoff, -6 * \kspacedy + \kspaceyoff) {\includegraphics[width=\kspacewidth]{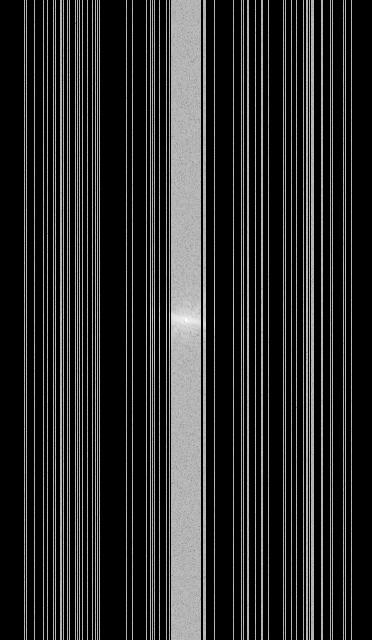}};
		\foreach \i in {1, 2, 3}
		{
			\pgfmathsetmacro{\fr}{\i/4}
			\node at ($(kspace3.south west)!\fr!(kspace6.south west)$) {\( \cdot \)};
			\node at ($(kspace3.north east)!\fr!(kspace6.north east)$) {\( \cdot \)};
		}
		\foreach [count=\i] \which in {00, 20, 40, 90}
		{
			\node[inner sep=0, outer sep=0] (reconstruction\i) at (1cm + \i * \kspacewidth + \i * \ppad, 0.3) {%
				\includegraphics[angle=180,origin=c,width=\kspacewidth]{\prefix/0\which/reconstruction_unnormalized.png}%
			};
			\node at ($(reconstruction\i.north) + (0, 0.2cm)$) {%
				\ifthenelse{\i=1}{\( u^{(0)} \in \R^n_{\geq0} \)}{%
					\ifthenelse{\i<4}{\( u^{(\which)} \)}{\( u^{(K)} \eqqcolon \optimal{u} \)}
				}%
			};
			\node at (1cm + \i * \kspacewidth + \i * \ppad, -2.5) {%
				\ifthenelse{\i=1}{\( \Sigma^{(0)} \)}{%
					\ifthenelse{\i<4}{\( \Sigma^{(\which)} \)}{\( \Sigma^{(K)} \eqqcolon \Sigma^* \)}
				}
			};
			\node at (0.9cm + \i * \kspacewidth + \i * \ppad, -1.95) {\( \ddots \)};
			\foreach [count=\isigma] \sigmawhich in {00, 01, 02, 14}
			{
				\ifthenelse{\isigma<4}{\def\offi{\isigma}}{\def\offi{6}}
				\node [inner sep=0, outer sep=0] (sensitivity maps\i\isigma) at (0.7cm + \i * \kspacewidth + \i * \ppad + \offi * \sensedx, -\offi * \sensedy - 1.0cm) {\includegraphics[width=\sensewidth]{\prefix/0\which/coil_\sigmawhich.png}};
			}
		}
		\node[rectangle, draw, minimum width=0.6cm, minimum height=0.6cm] (initial guess box) at ($(reconstruction1.west) + (-1, 0)$) {\eqref{eq:initial guess}};
		\draw [latex-] (initial guess box) -- ++(-1, 0);
		\draw [-latex] (initial guess box) -- (reconstruction1);
		\draw [-latex] (initial guess box) |- (sensitivity maps14.west);
		\foreach \i in {1, 2, 3}
		{
			\pgfmathtruncatemacro{\j}{\i + 1}
			\node [inner sep=0,outer sep=0] (mid\i) at ($(reconstruction\i)!.5!(reconstruction\j)$) {\( \square \)};
			\draw (reconstruction\i) -- (mid\i) [-latex] -- (reconstruction\j);
			\draw [latex-,dashed] (mid\i) -- ++(0, 0.5) node [above] {\( z \)};
		}

		\begin{scope}[on background layer]
			\draw [thick, draw=green!20!black, opacity=0.5, fill=green!50!black] ($(reconstruction1.north west)+(-0.3, 0.5)$) rectangle ($(reconstruction4.south east) + (+0.3, -1.5)$);
		\end{scope}
		\begin{scope}[shift={(16, 1.6)}]
			\node (upper) [draw, rounded corners] at (0, 0) {\( \bigl( \, \cdot \, - L^{-1}_u \nabla H(\,\cdot\,, \Sigma^{(k)}) \bigr) \)};
			\draw [->] (-1.1, 0.8) -- (upper.north -| -1.1, 2);
			\node at (-1.3, 1.) {\( u^{(k)} \)};
			\node (proxU) [draw, rounded corners] at (0, -1) {\( u^{(k+1)} = (\, \cdot \,)_+ \)};
			\node (nablaS) [draw, rounded corners] at (0, -2.3) {\( \bigl( \, \cdot \, - L^{-1}_\Sigma \nabla H(u^{(k+1)}, \, \cdot \,)\bigr) \)};
			\draw (1.1, 0.8) -- ($(upper.north -| 1.1, 0.8) + (0.,0.2)$);
			\draw [loosely dashed] ($(upper.north -| 1.1, 0.8) + (0.,0.15)$) -- ++(0, -2.1);
			\draw [->] ($(upper.north -| 1.1, 0.8) + (0.,-1.95)$) -- (nablaS.north -| 1.1,0);
			\node at (1.1, 1) {\( \Sigma^{(k)} \)};
			\node (lower) [draw, rounded corners] at (0, -3.3) {\( \mathcal{S}^{-1} \bigl( \operatorname{diag}(\xi_i + \mu)^{-1} \mathcal S(\mu\,\cdot\,) \bigr) \)};
			\begin{scope}[on background layer]
				\draw [thick, draw=black, fill opacity=0.5, fill=green!50!black] ($(upper.north west -| lower.south west)+(-0.3, 0.3)$) rectangle ($(lower.east |- lower.south)+(0.3, -0.3)$);
			\end{scope}
			\begin{scope}[on background layer]
				\draw [thick, draw=blue!20!black, opacity=0.5, fill=blue!30!black] ($(upper.north west -| lower.south west)+(-0.15, 0.15)$) rectangle ($(lower.east |- proxU.south)+(0.15, -0.15)$);
			\end{scope}
			\begin{scope}[on background layer]
				\draw [thick, draw=red!20!black, opacity=0.5, fill=red!30!black] ($(nablaS.north west -| lower.west)+(-0.15, 0.15)$) rectangle ($(lower.east |- lower.south)+(0.15, -0.15)$);
			\end{scope}
			\draw [<-, dashed] (upper.east) -- (2.5, 0) node [right] {\( z \)};
			\draw [<-, dashed] (nablaS.east) -- (2.5, 0 |- nablaS.east) node [right] {\( z \)};
			\draw [->] (upper) -- (proxU);
			\draw [->] (nablaS) -- (lower);
			\draw (proxU) -| (-1.3, -1.7);
			\draw [loosely dashed] (-1.3, -1.8) -- ++(0, -2);
			\draw [->] (-1.3, -3.9) -- (-1.3, -4.2) node [below] {\( u^{(k + 1)} \)};
			\draw [->] (lower.south -| 1.1, 0) -- (1.1, -4.2) node [below] {\( \Sigma^{(k + 1)} \)};
		\end{scope}
		\draw [dashed, opacity=0.5] (mid3.north west) to [in=-110,out=20] ($(upper.north west -| lower.south west)+(-0.3, 0.3)$);
		\draw [dashed, opacity=0.5] (mid3.south west) to [in=110,out=-20] ($(lower.south west |- lower.south)+(-0.3, -0.3)$);
	\end{tikzpicture}}%
	\caption{%
		\added{%
			Sketch of the reconstruction algorithm:
			To jointly reconstruct the spin density \( u \) and the sensitivity maps \( \Sigma \), we impose data-fidelity, image-regularity, and coil-regularity in the iterations of \acrshort{ipalm}~\cite{pock_inertial_2016}.
			The function \( H \) incorporates our learned regularizer acting on \( u \).
			Details are discussed in~\cref{ssec:mri}.
		}
	}
	\label{fig:reco}
\end{figure*}
Data-driven methods typically require large training datasets.
In particular, state-of-the-art networks mimicking iterative schemes like the end-to-end \gls{vn}~\cite{sriram_endtoend_2020} as well as AUTOMAP~\cite{zhu_image_2018} (and related methods such as~\cite{zhou2020dudornet}) require image-data pairs for training.
Such image-data pairs are scarce in medical applications, where\added{as} reconstructed images are much more abundantly available\added{ in the form of DICOM data~\cite{zbontar_fastmri_2018}}.
Data-driven methods that only assume access to reference images (but still allow for some kind of data-\replaced{fidelity}{consistency}) include \glspl{gan}~\cite{Narnhofer2019} and score-based diffusion models~\cite{chung_scoremri_2022}.
Both methods share the problem that it is not obvious how the encoded prior information is best used:
\Glspl{gan} suffer from the range-dilemma~\cite{bora_compressed_2017}
\replaced{%
	and authors have proposed to optimize the parameters of the \gls{gan} at inference time~\cite{Narnhofer2019}, effectively turning it into a deep image prior~\cite{ulyanov_dip_2018}.
	Diffusion models have shown remarkable results in \gls{mri} reconstruction~\cite{chung_scoremri_2022,song2022solving,luo_bayesian_2023}, but it is still an open question how to optimally incorporate data-fidelity in the reverse diffusion process (see~\cite{feng2023scorebased} and the discussion in~\cref{ssec:related work} for an overview of proposed methods).
}{%
	, and for score-based diffusion models it is often not clear which distribution the reconstructions are drawn from.
}
In addition, (with the exception of score-based diffusion models) \emph{all} of the mentioned models effectively act as a point estimator, mapping a \( k \)-space datum to an image.
However, having access to a distribution of reconstructions and, consequently, being able to quantify uncertainty in the reconstruction is of utmost importance in medical applications.
Further, these point estimators typically assume a particular acquisition modality and do not generalize with respect to acquisition masks~\cite{chung_scoremri_2022}.
In fact, it has been shown that reconstruction quality of some data-driven approaches deteriorates when more data becomes available~\cite{Antun2020}.
This severely hampers adoption of these methods in clinical practice.
In any case, a method that marries the great representation power of data-driven deep neural networks with the interpretability of variational approaches is still sought after.

In this work we pursue a principled approach to parallel \gls{mri} that combines the merits of modern data-driven approaches with the benefits of classical variational approaches.
In particular, we learn a highly expressive image-space prior in the well-known maximum-likelihood framework.
In contrast to most other data-driven approaches, training our model does not require image-data pairs, but only assumes access to a database of reference reconstructions.
To solve different \gls{pi} reconstruction problems, we propose a novel algorithm to jointly estimate the image as well as the sensitivity maps in a \gls{sense}-type~\cite{pruessmann} forward model.
\added{%
	A sketch of our proposed approach is shown in~\cref{fig:reco}.
}
We demonstrate that combining this prior with suitable data-likelihood terms yields competitive performance for a multitude of \gls{pi} problems.
Specifically, the variational formulation makes our method agnostic to \replaced{sub-sampling}{acquisition} patterns.
Additionally, the accompanying probabilistic interpretation naturally enables experts to explore the full posterior distribution of any reconstruction problem.
In particular, we analyze the posterior expectation as well as the pixel-wise marginal posterior variance.

\subsection{On the Hunt for the Optimal Regularizer}
The classical variational approach for solving inverse problems amounts to minimizing an energy functional
\begin{equation}
	\min_x E(x, z) \coloneqq D(x, z) + R(x).%
	\label{eq:variational}
\end{equation}
Here, \( x \) is the reconstruction given the datum \( z \).
The data-fidelity term \( D \) models the physical acquisition process, including stochasticity.
For example, if \( A \) models the physical acquisition process, \( D(x, z) = \sfrac{1}{2}\norm{Ax - z}_2^2 \) is optimal under the assumption of Gaussian measurement noise.
On the other hand, the regularization term \( R \) encodes prior knowledge about the solution.
Classical regularizers \( R \) include magnitude penalization (e.g.\ the Tikhonov functional \( x \mapsto \norm{x}_2^2 \)) or encode smoothness assumptions (e.g.\ the \gls{tv} \( x \mapsto \sum_{i=1}^n \sqrt{{(\mathrm{D} x)}_{i}^2 + {(\mathrm{D} x)}_{i + n}^2} \)).

Often, \cref{eq:variational} is used without fully acknowledging its probabilistic interpretation:
It computes the \gls{map} estimate of the posterior \( p(x\mid z) \propto p(z \mid x) p_R(x) \) composed of a likelihood \( -\log p(z \mid x) \propto D(x, z) \) and a prior \( -\log p_R(x) \propto R(x) \).
Hand-crafted regularizers, in general, can not faithfully represent the intricate statistics of the underlying distribution.
Since these regularizers are overly simplistic, methods like early stopping the optimization of~\cref{eq:variational} have emerged~\cite{kobler_total_2020}.
Clearly, the need for such heuristics arises only because hand-crafted regularizers are a bad model for natural or medical images.

In this paper we pose the following question:
Can we learn a regularizer such that \( p_R \) is indistinguishable from the underlying reference distribution?
We answer this by encoding the regularizer as a neural network endowed with learnable parameters, which we train on reference data using maximum likelihood (see~\cref{sec:methods}).
\added{%
	This method of learning an unnormalized probability density is known as learning an \gls{ebm}~\cite{du_implicit_2019,nijkamp_anatomy_2019}, and has recently been used in the context of \gls{mri} reconstruction by~\cite{roth_fields_2005,Guan_energy_2022,tu_collaborative_2023}.
}
After training, the network can be used in arbitrary reconstruction tasks by embedding it in~\cref{eq:variational}, thereby combining the versatility of variational approaches with the great representation power of deep neural networks.

On one hand, concerning versatility, the strict separation of data likelihood and prior makes this approach agnostic to the \replaced{%
	number of receiver coils and sub-sampling pattern.%
}{%
	acquisition protocol.
	We can reconstruct images from single- and multi-coil data with arbitrary acquisition masks.%
}
Further, the probabilistic interpretation enables access to a distribution of reconstructions \( p(x \mid z) \), and we can compute different Bayesian estimators.
In addition to the \gls{map} estimate (\( \delta_x \) is the Dirac measure centered at \( x \))
\begin{equation}
    \max_x \int p(x^\prime \mid z)\ \mathrm{d}\delta_x(x^\prime),
\end{equation}
we exploit the distribution to compute the \gls{mmse} estimate
\begin{equation}
    \min_x \int  p(x^\prime \mid z)\norm{x - x^\prime}^2 \ \mathrm{d}x^\prime \eqqcolon \mathbb{E}[x \mid z],%
    \label{eq:mmse}
\end{equation}
as well as the pixel-wise marginal variance (at the \( i^\text{th} \) pixel)
\begin{equation}
	\int {\big( x^\prime_i - \mathbb{E}{[x \mid z]}_i \big)}^2\ p(x^\prime_i\mid z)\ \mathrm{d}x^\prime_i \eqqcolon (\operatorname{Var}[x\mid z])_i.%
    \label{eq:variance}
\end{equation}

On the other hand, the representation power is empirically demonstrated by two experiments:
First, our model is capable of synthesizing realistic images \emph{without} any data.
Second, we show that we can faithfully reconstruct images from severely ill-posed problems, e.g.\ when using random acquisition masks, where profound knowledge of the underlying anatomy is required.
\subsection{Parallel \texorpdfstring{\gls{mri}}{MRI}}
\label{ssec:parallel imaging intro}
In modern parallel \gls{mri} systems, the view of~\cref{eq:variational} is overly simplistic:
These systems typically utilize coil arrays with spatially varying sensitivity maps~\cite{pruessmann} which are, in general, entirely unknown.
In other words, the physical acquisition model is not well specified.
Thus, for any real practical reconstruction problem, a great challenge lies in estimating these sensitivity maps accurately.

Many strategies for estimating the sensitivity maps in an off-line step have been proposed (e.g. ESPIRiT~\cite{uecker_espirit_13,McKenzie2002}).
Typically, these require the acquisition of fully sampled \gls{acl} in the \( k \)-space center such that, in essence, a low-resolution reconstruction along with the corresponding sensitivity maps can be estimated.
The disadvantage of such methods is two-fold:
First, assuming a non-Cartesian \replaced{sub-sampling pattern}{acquisition protocol}, the acquisition of the \gls{acl} essentially requires an additional (albeit low-resolution) scan.
This increases examination time and opens the possibility of patient movement and subsequent misalignment artifacts.
Second, high-frequency regions of the \( k \)-space are not taken into account for estimating the sensitivity maps.
Generally, the sensitivity maps can be assumed to be smooth (indeed we also make this assumption).
However, since errors in the sensitivity estimation adversely affect downstream tasks significantly, it is vital to exploit all available data to the best extent.

In this work, similar to~\cite{Knoll2011,uecker_image_2008,Ying2007}, we propose to estimate the image and sensitivity maps jointly.
This joint estimation hinges on the observation that the sensitivity maps are much smoother than the imaged anatomy.
To enforce this smoothness and simultaneously resolve ambiguities, we impose a simple quadratic penalization on the spatial gradient of the sensitivity maps.
The joint estimation allows us to utilize all available \( k \)-space data to estimate the image as well as the sensitivity maps.
The resulting optimization problem can be efficiently solved by \replaced{accelerated}{momentum-based} non-convex optimization algorithms.
Reconstructing a parallel \gls{mri} image takes about \qty{5}{\second} on consumer hardware.
\subsection{Related Work}
\label{ssec:related work}
\added{%
	In this section we review some related work on learned \gls{mri} reconstruction, but restrict our attention to methods that share similarities with our proposed approach.
	For a more general overview of data-driven \gls{mri} reconstruction we refer to~\cite{Zeng2021}.
	The authors of~\cite{Antun2020} give a very broad overview of potential risks of using modern techniques for medical image reconstruction in general.
}
\subsubsection{Energy-based Models}
\added{%
	After submission of this paper, we were made aware of concurrent works that also utilize \glspl{ebm} for \gls{mri} reconstruction.
	In~\cite{Guan_energy_2022}, the authors propose to learn a regularizer and use proximal gradient descent for inference.
	The main difference to our approach is that for parallel \gls{mri}, their algorithm assumes access to sensitivity maps, which have to be precomputed using, e.g., ESPIRiT~\cite{uecker_espirit_13}.
	Additionally, their algorithm requires hand-tuning of step-sizes and is relatively slow.
	On the contrary, we propose an algorithm for joint reconstruction of image and coil sensitivities which does not require hand-tuning of step-sizes, and is fast due to being accelerated.%
}

\added{%
	The authors extended their work in~\cite{tu_collaborative_2023}, where they learn an \gls{ebm} for both image- and k-space.
	While this obviates the need for sensitivity estimation, it requires to train two independent networks that need to be balanced at inference time.
	Additionally, the k-space \gls{ebm} requires fully-sampled reference k-space data, which is scarcely available.
	In contrast, we only require reference DICOM (magnitude) images to train one network.
	In addition, we perform a data-independent analysis of the learned regularizer as well as uncertainty analysis.%
}
\subsubsection{Diffusion Models}
\added{%
	Diffusion models~\cite{song2021scorebased,song2022solving,luo_bayesian_2023,ajil_robust_comporessed_2021} aim to model the gradient of the log-prior while undergoing a diffusion process:
	Let \( p_t \) denote the data distribution at diffusion time \( t \).
	The aim is to learn a time-conditional score network \( s_\theta \) such that \( s_\theta(\,\cdot\,, t) \approx \nabla \log p_t \) for all \( t > 0 \).
	This approach shares many similarities with \glspl{ebm}, where indeed it should hold that \( \nabla R = s_\theta(\,\cdot\,, 0) \).%
}

\added{%
	Diffusion models can be used to generate unconditional samples from the data distribution~\cite{song2021scorebased}.
	This is computationally demanding as it requires solving a \gls{sde} with high accuracy, which typically needs thousand of gradient evaluations~\cite{chung_scoremri_2022}.
	For inverse problems, it is not clear how to optimally incorporate data-fidelity into the \gls{sde}.
	Proposed approaches include data projection~\cite{song2022solving}, annealed Langevin dynamics~\cite{ajil_robust_comporessed_2021} and diffusion posterior sampling~\cite{chung2023diffusion}.
	These approaches require hand-tuning of parameters, sometimes at each step of the reverse diffusion~\cite{ajil_robust_comporessed_2021,chung2023diffusion}.
	The recent work of~\cite{feng2023scorebased} showed that none of these methods generate samples from the true posterior distribution and propose to augment the score models with normalizing flows.
	While their inference algorithm is parameter-free, the approach is opaque due to the introduction of the normalizing flow and is still computationally demanding.%
}

\added{%
	In contrast, \glspl{ebm} enjoy a natural probabilistic interpretation with access to an analytic expression of the posterior.
	In addition, \gls{map} inference does not require solving an \gls{sde}, but can be done by efficient optimization algorithms.
	Having access to the function value (as opposed to only the gradient) also has practical applications, such as being able to inspect the regularization landscape (see~\cref{fig:pdf sampling}) and utilizing backtracking in optimization algorithms~\cite{pock_inertial_2016}.%
}

\added{%
	In the works of~\cite{luo_bayesian_2023,ajil_robust_comporessed_2021}, parallel imaging is tackled by off-line sensitivity estimation, which comes with the drawbacks outlined in~\cref{ssec:parallel imaging intro}.
	The authors of~\cite{chung_scoremri_2022} propose to reconstruct individual coil images using a model trained solely on \gls{rss} reconstructions.
	While the results are impressive, the computational cost consequently depends on the number of coils utilized in the physical scanner, and the authors report reconstruction times of up to \qty{10}{\minute}.
	In our joint reconstruction algorithm, the gradient of the network is only evaluated once per iteration, irrespective of the number of coils.
	Imposing spatial regularity on the coils is extremely fast, where at each iteration we only have to compute very few fast Fourier transforms (see~\cref{ssec:mri}).%
}

\subsubsection{Joint reconstruction}
\added{%
	The common approach to parallel imaging is based on a two-step approach:
	Coil sensitivities are estimated (typically from \gls{acl} regions) in an off-line step, after which the image is reconstructed.
	A joint reconstruction was first proposed by~\cite{Ying2007} based on alternating minimization, where they explicitly parametrize the sensitivities with low-order polynomials.
	In~\cite{uecker_image_2008}, the authors propose an iteratively regularized Gauss-Newton algorithm that enforces spatial smoothness of the sensitivity maps during the iterations.
	Their algorithm is not guaranteed to converge for arbitrary initializations, requires hand-tuning of parameters at each update step and leads to noisy solutions when run for too long.
	This algorithm was extended to incorporate classical variational penalties, such as the \gls{tv}~\cite{rudin_nonlinear_1992} or the \gls{tgv}~\cite{bredies_total_2010}, by the authors of~\cite{Knoll2011}, in order to suppress noisy solutions.%
}

\added{%
	Indeed, their approach shares a lot of similarities with our proposed approach.
	The differences can be summarized as follows:
	Instead of hand-crafted regularizers, we employ modern generative learning techniques to learn an expressive regularizer from data, which leads to state-of-the-art reconstructions.
	In addition, instead of the iteratively regularized Gauss-Newton algorithm with additional non-trivial sub-problems for the variational penalties, we employ the \gls{ipalm} algorithm~\cite{pock_inertial_2016} for optimization.
	Thus, we can guarantee convergence and only require hand-tuning of two parameters (see~\cref{ssec:mri}).
	A sketch of our reconstruction algorithm is shown in~\cref{fig:reco}.
}

\added{%
	For completeness, we mention that the end-to-end variational network of~\cite{sriram_endtoend_2020} also estimates the sensitivities jointly with the image.
	However, their approach differs significantly to ours in that they learn a mapping from \( k \)-space to image-space \emph{discriminatively}, where they utilize the estimated sensitivity maps to enforce data fidelity.
	Thus, the network only works well with a particular sub-sampling pattern and coil configuration.
	In contrast, we learn image features \emph{generatively} and impose hand-crafted spatial regularity onto the sensitivity maps, therefore being agnostic to the sub-sampling pattern as well as coil configuration.%
}

\section{Methods}%
\label{sec:methods}
\subsection{Maximum-likelihood training}%
\label{ssec:methods ml}
To learn a regularizer such that its induced distribution is indistinguishable from the underlying reference distribution we proceed as follows:
We equip the regularizer with parameters and denote with \( \{ R_\theta : \theta \in \Theta \} \) the family of \( \theta \)-parametrized functions \( \Function{R_\theta}{\set{X}}{\R} \).
\( \Theta \) is a suitably selected set of parameters, \( \set{X} \) is the space of the underlying distribution.
We discuss our particular choice of the \( \theta \)-parametrized family in~\cref{ssec:details}.
\( R_\theta \) induces a Gibbs distribution on \( \set{X} \) with density
\begin{equation}
	p_\theta(x) \coloneqq \exp(-R_\theta(x))Z^{-1}_\theta,%
	\label{eq:gibbs}
\end{equation}
where \( Z_\theta = \int_\set{X}\exp(-R_\theta(\xi))\,\mathrm{d}\xi \) is the partition function.
We find the optimal parameters by maximizing the likelihood of reference data --- drawn from a reference distribution \( p_\text{data} \) --- under our model:
\begin{equation}
	\min_{\theta \in \Theta} \left\{ \Gamma(\theta) \coloneqq \mathbb{E}_{x\sim p_{\text{data}}}[-\log p_{\theta}(x)] \right\}.%
	\label{eq:ml}
\end{equation}
The loss function~\cref{eq:ml} admits the gradient~\cite{hinton_training_2002} \( \nabla \Gamma(\theta) = \mathbb{E}_{x\sim p_\text{data}} [\nabla_\theta R_\theta(x)] - \mathbb{E}_{x\sim p_\theta} [\nabla_\theta R_\theta(x)] \).

For any interesting regularizer, computing the partition function \( Z_\theta \) is intractable.
Thus, we resort to \gls{mcmc} techniques to approximate \( \mathbb{E}_{x \sim p_\theta} [\nabla_\theta R_\theta(x)] \).
In particular, following~\cite{du_implicit_2019} we use the \gls{ula}~\cite{roberts_exponential_1996}, which iterates
\begin{equation}
	x^{(j)} \sim \mathcal{N}\Bigl(x^{(j - 1)} + \frac{\zeta}{2}\nabla \log p_\theta (x^{(j - 1)}), \zeta \mathrm{Id}\Bigr),\ j \in \llbracket 1, J \rrbracket.
    \label{eq:langevin}
\end{equation}
Here, \( \mathcal{N}(\mu, \Sigma) \) denotes the normal distribution on \( \set{X} \) with mean \( \mu \) and covariance \( \Sigma \), and \( \llbracket \,\cdot\,,\,\cdot\,\rrbracket \) denotes an integer interval.
Further, \( \zeta > 0 \) is the discretization time step of the associated continuous-time Langevin diffusion \acrlong{sde}, which is known to be \( p_\theta \)-stationary~\cite{roberts_exponential_1996,roberts_optimal_1998}.
The time discretization biases \gls{ula} asymptotically, which could be corrected via metropolization.
However, the poor non-asymptotic performance of metropolized \gls{ula}~\cite{durmus_efficient_18,bortoli_langevin_21} makes this less useful in practice, and we do not use metropolization.
Note that by~\cref{eq:gibbs} \( \nabla \log p_\theta = -\nabla R_\theta \).

\added{%
	In general, maximum-likelihood based learning of \glspl{ebm} is known to be unstable~\cite{du_improved_2020,nijkamp_anatomy_2019} due to the slow convergence of \gls{ula}~\eqref{eq:langevin}.
	To minimize the burn-in time and aid convergence of our sampler, we utilize persistent initialization~\cite{zhu_filters_1998,tieleman_persistent_2008}:
	Assuming \( p_\theta \) changes only slightly during each update to \( \theta \), samples from previous learning states are good initial guesses for the current iteration.
	We implement this idea using a replay buffer that holds samples from previous iterates.
	After iterating~\eqref{eq:langevin} starting from \( x^{(0)} \) drawn from the replay buffer, we write \( x^{(J)} \) back into the buffer with \( 1 - \pi_{\text{reinit}} \) chance.
	Otherwise, we draw a random sample from \( p_{\text{data}} \) to write it to the buffer.
	Upon reinitialization, to help mode coverage, we randomly permute the pixels of (on average) every second sample.
	We refer to~\cite{hinton_training_2002,du_implicit_2019,nijkamp_anatomy_2019} for more information of maximum-likelihood training of \glspl{ebm}.
}
\subsection{Reconstruction algorithm}%
\label{ssec:mri}
We utilize an image-space SENSE-type~\cite{pruessmann} forward model for parallel \gls{mri}.
In detail, we relate the real-valued spin density \( x \in \R^n_{\geq 0} \) to the noisy measurement data \( z = {(z_1,\dotsc,z_C)}^\top \in \mathbb{C}^{m} \).
Here, \( m = CK \) combines the \( K \) acquired \( k \)-space data points (not necessarily on a Cartesian grid) of \( C \in \mathbb{N} \) receiver coils.
The forward operator
\begin{equation}
	 \Function{A}{\R^n}{\mathbb{C}^m}, \quad x \mapsto \begin{pmatrix}
		\mathcal{F}_M(\sigma_1 \odot x) \\
		\mathcal{F}_M(\sigma_2 \odot x) \\
		\vdotswithin{\mathcal{F}_M(\sigma_1 \odot x)} \\
		\mathcal{F}_M(\sigma_{C} \odot x)
	\end{pmatrix}
\end{equation}
utilizes the sensitivity maps \( {(\sigma_c)}_{c=1}^{C} \in {(\mathbb{C}^n)}^{C} \) of the \( C \) receiver coils, as well as the sampling operator \( \Function{\mathcal{F}_M}{\mathbb{C}^n}{\mathbb{C}^{K}} \) defined by a \( k \)-space trajectory \( M \).
\added{%
	In our model, we assume a real-valued spin-density and empirically demonstrate good performance on the fastMRI dataset~\cite{zbontar_fastmri_2018} in~\cref{ssec:parallel imaging}.
	In phase-sensitive imaging however, a complex spin-density needs to be assumed.
	Our approach can be generalized to this setting by, e.g. splitting real and imaginary channels as in~\cite{Narnhofer2019,sriram_endtoend_2020}, and training the regularizer on this data.
}

To ease notation, we define \( \Sigma \coloneqq {(\sigma_c)}_{c=1}^C \), \( \mathcal{C} \coloneqq {(\mathbb{C}^n)}^C \), and \( |\,\cdot\,|_{\mathcal{C}} \colon \mathcal{C} \to \R^n_{\geq 0} \), \( {(\sigma_c)}_{c=1}^C \mapsto \sqrt{\sum_{c=1}^C |\sigma_c|^2} \). (\( |\,\cdot\,| \) is the complex modulus acting element-wise on its argument.)
In the case of sub-sampling on a Cartesian grid, \( \mathcal{F}_M \) encodes the Fourier transform followed by the multiplication with a binary mask.
In such \gls{sense}-type setups, to quantitatively compare to fully-sampled \gls{rss} reconstructions, the spin density \( x \) has to be re-weighted by \( |\Sigma|_{\mathcal{C}} \)~\cite{uecker_image_2008}:
Denoting the reconstructed image \( u \in \R^n_{\geq 0} \), it is given by
\( 
	u = x \odot |\Sigma|_{\mathcal{C}},
\)
which motivates an immediate change of variables 
\begin{equation}
    	x = u \oslash |\Sigma|_{\mathcal{C}}.
    	\label{eq:change of variables}
\end{equation}

Instead of estimating the spin density (and the sensitivity maps), we propose to directly estimate the image \( u \) (and the sensitivity maps).
Formally, given the noisy measurement data \( z = {(z_1,\dotsc, z_C )}^{\top} \)  we propose to find
\begin{equation}
	\begin{aligned}
		\argmin_{(u, \Sigma) \in \R^n \times \mathcal{C}} H(u, \Sigma) + \delta_{\mathbb{R}_{\geq 0}^n}(u) + \mu F(\Sigma)
	\end{aligned}
	\label{eq:mri problem}
\end{equation}
where
\begin{equation}
	\begin{aligned}
		\Function{H}{\R^n \times \mathcal{C}&}{\R}, \\
		(u, {(\sigma_c)}_{c=1}^C) &\mapsto \frac{1}{2} \norm{%
			\begin{pmatrix}
				\mathcal{F}_M(\sigma_1 \odot u \oslash |\Sigma|_{\mathcal{C}}) \\
				\vdotswithin{%
					\mathcal{F}_M(\sigma_1 \dot u \slash |\Sigma|_{\mathcal{C}})%
				} \\
				\mathcal{F}_M(\sigma_{C} \odot u \oslash |\Sigma|_{\mathcal{C}})
			\end{pmatrix} - \begin{pmatrix}
				z_1 \\ \vdotswithin{y_1} \\ z_C
			\end{pmatrix}%
		}_2^2\\
		&\phantom{\mapsto}\ + \lambda R (u)
	\end{aligned}%
	\label{eq:H}
\end{equation}
combines \deleted{the} data fidelity and image-regularization and 
\begin{equation}
	\begin{aligned}
		\Function{F}{\mathcal{C}&}{\R_{\geq 0}}, \\
		{(\sigma_c)}_{c=1}^C &\mapsto \frac{1}{2}\sum_{c=1}^{C} \big( \norm{\mathrm{D} \mathfrak{Re}(\sigma_c)}_2^2 + \norm{\mathrm{D} \mathfrak{Im}(\sigma_c)}_2^2 \big)
	\end{aligned}
\end{equation}
encodes the smoothness prior on the sensitivity maps.
In the above, \( \delta_{\R^n_{\geq 0}} \) enforces non-negativity on the spin\replaced{ }{-}density using the indicator function
\begin{equation}
	\Function{\delta_{\set{A}}}{\R^n}{\{ 0, \infty \}}, \quad x \mapsto \begin{cases} 0 & \text{ if } x \in \set{A}, \\
	\infty & \text{ else.}\end{cases}
\end{equation}
\( \Function{\mathrm{D}}{\R^n}{\R^{2n}} \) is the discrete gradient operator such that \( (\mathrm{D} x) \) contains vertically stacked horizontal and vertical first-order finite differences (see, e.g.,~\cite{chambolle_introduction_2016}).
It implements physically motivated Dirichlet boundary conditions (\( \sigma_c = 0 \) outside of the image domain \( \forall c \)).
\( \lambda \in \R_+ \) and \( \mu \in \R_+ \) are scalars trading off the strength of the regularization on the image and the sensitivity maps respectively.

Observe that the division by \( |\Sigma|_{\mathcal{C}} \) in~\cref{eq:H} follows from the identification~\cref{eq:change of variables}.
This formulation has the advantage that the sensitivity maps \( {(\sigma_c)}_{c=1}^C \) are implicitly normalized, and we do not have to impose a normalization constraint explicitly during optimization.
Further, the existence of a minimizer is ensured by coercivity of~\cref{eq:mri problem}, which is a direct result of the enforced Dirichlet boundary conditions.

We solve~\cref{eq:mri problem} using the \gls{ipalm} algorithm~\cite{pock_inertial_2016} with Lipschitz backtracking, summarized in~\cref{alg:ipalm}.
\added{%
	The algorithm utilizes the combined fidelity-regularity functional \( H \) defined in~\eqref{eq:H}.
	For our learned regularizer, we would have \( R = R_\theta \) as in~\eqref{eq:reg}, whereas for a hand crafted regularizer \( R \) may be TV~\eqref{eq:charb}.
}
All gradients are understood in the \( \mathbb{C}\mathbb{R} \)-sense~\cite{kreutz_complex_2009}.
Recall that the proximal operator \( \Function{\prox_{\alpha G}}{\set{H}}{\set{H}} \) of a proper extended real-valued function \( \Function{G}{\mathcal{H}}{\interval[open left]{-\infty}{\infty}} \) (\( \mathcal{H} \) is a Hilbert space) is the map \( \bar{x} \mapsto \argmin_x \sfrac{1}{2} \norm{\bar{x} - x}^2 + \alpha G(x) \).
Observe that \( \prox_{\delta_{\R^n_+}} \) retrieves the positive part of its argument, that is \( \prox_{\delta_{\R^n_+}}(x) = {(x)}_+ \).
\( \prox_{\mu F} \) can be solved in closed form by utilizing the discrete sine transform (see~\cite[Chapter 19.4]{press_numerical_1992} for a more rigorous discussion) as (denoting \( \hat{\imath} \coloneqq \sqrt{-1} \))
\begin{equation}
	{(\sigma_c)}_{c=1}^C \mapsto {\big(Q_\mu(\mathfrak{Re}(\sigma_c)) + \hat{\imath} Q_\mu(\mathfrak{Im}(\sigma_c))\big)}_{i=1}^C.
\end{equation}
Here, \( Q_\mu \colon y \mapsto \mathcal{S}^{-1}\bigl({\operatorname{diag}(\xi_i + \mu)}^{-1}\mathcal{S}(\mu y) \bigr) \) uses the discrete sine transform \( \mathcal{S} \) and the eigenvalues \( \xi_i \) of the discrete Laplace operator, which are of the form \( \xi_i = 2 - 2 \cos \phi_i \) for equally spaced angles \( \phi_i \).
We initialize the algorithm with the \gls{zf} \gls{rss} reconstruction and the corresponding sensitivity maps (for all \( c \in \llbracket 1, C \rrbracket \))
\begin{equation}
	u^{(0)} = \sqrt{\sum_{c=1}^C|\mathcal{F}_M^*(z_c)|^2}\ \text{ and }\ \sigma_c^{(0)} = \mathcal{F}_M^*(z_c) \oslash u^{(0)}.%
	\label{eq:initial guess}
\end{equation}
\added{A sketch of the reconstruction algorithm is shown in~\cref{fig:reco}.}
\begin{algorithm}
	\DontPrintSemicolon%
	\SetKwInOut{Input}{Input}
	\SetKwInOut{Output}{Output}
	\Input{\( (u^{(0)}, \Sigma^{(0)}) \in \R^n \times \mathcal{C} \), \( K \in \mathbb{N} \), \( \gamma_1 \in (0, 1) \), \( \gamma_2 \in (0, 1) \), initial \( (L_x, L_\Sigma) \in \R_+ \times \R_+ \)}
	\Output{\( (u^{(K)}, \Sigma^{(K)}) \) solving~\cref{eq:mri problem}}
	\( (u^{(1)}, \Sigma^{(1)}) = (u^{(0)}, \Sigma^{(0)}) \)\;
	\For{\( k \in \llbracket 1, K - 1 \rrbracket \)}{
		\( \bar{u} = u^{(k)} + \frac{k}{k + 3} (u^{(k)} - u^{(k - 1)})\)\;
		\( (u^{(k + 1)}, L_x) = \operatorname{bt}(H(\,\cdot\,,\Sigma^{(k)}), \delta_{\R^n_+}, \bar{u}, L_u, \gamma_1, \gamma_2) \)\;
		\( \bar{\Sigma} = \Sigma^{(k)} + \frac{k}{k + 3} (\Sigma^{(k)} - \Sigma^{(k - 1)}) \)\;
		\( (\Sigma^{(k+1)}, L_\Sigma) = \operatorname{bt}(H(u^{(k + 1)},\,\cdot\,), \mu F, \bar{\Sigma}, L_\Sigma, \gamma_1, \gamma_2) \)\;
	}
	\caption{%
		\gls{ipalm}~\cite{pock_inertial_2016} instantiation to solve \eqref{eq:mri problem}.
		\( \operatorname{bt} \) is~\cref{alg:backtrack}.
	}%
	\label{alg:ipalm}
\end{algorithm}
\begin{algorithm}
	\DontPrintSemicolon%
	\SetKwInOut{Input}{Input}
	\SetKwInOut{Output}{Output}
	\Input{\( E \), \( P \), \( x_0 \), \( L_0 \), \( \gamma_1 \), \( \gamma_2 \) }
	\Output{\( (x, L) \)}
	\( L \leftarrow L_0 \)\;
	\For{ever}{%
		\( x = \prox_{L^{-1}P}(x_0 - L^{-1}\nabla E(x_0)) \)\;
		\( d = x - x_0 \)\;
		\uIf{
			\( E(x)  \leq E(x_0) + \langle \nabla E(x_0), d \rangle + \frac{L}{2} \norm{d}_2^2 \)
		}{%
			\( L \leftarrow \gamma_1 L \)\;
			\textbf{break}
		}
		\lElse{\( L \leftarrow L/\gamma_2 \)}
	}
	\caption{%
		Backtracking procedure to find the local Lipschitz constants in~\cref{alg:ipalm}.
	}%
	\label{alg:backtrack}
\end{algorithm}
\subsection{Experimental data}%
\label{ssec:experimental data}
For all experiments we utilize the fastMRI knee dataset~\cite{Knoll2020}.
Specifically, the training data are the \gls{rss} reconstructions of size \( \tilde{n} = \num{320} \times \num{320} \) of the multi-coil \gls{corpd} training split.
We used the central \num{11} slices to ensure reasonable training data, resulting in a total of \num{5324} training slices.
To have consistent intensity ranges during training, we normalized each slice by \( x \mapsto \sfrac{x - \min_i x_i}{\norm{x}_\infty - \min_i x_i} \) individually (this normalization was not performed for any of the reference methods).
For validation and testing, we used the multi-coil \gls{corpd} validation split, discarding samples with width \( w \notin \{ \num{368}, \num{372} \} \), leaving \num{91} scans.
The scans were split into \num{30} validation samples and \num{61} test samples by lexicographic ordering of the filenames.
To be consistent with training, we again restrict our interest to the central \num{11} slices, resulting in \num{330} validation slices and \num{671} test slices.
For the out-of-distribution experiments, we used the central \num{11} slices of the \gls{corpdfs} scans (again excluding width \( \notin \{ 368, 372 \} \)) in the fastMRI knee validation dataset.
\subsection{Network architecture and implementation details}%
\label{ssec:details}
The family of \( \theta \)-parametrized functions we consider in this \replaced{work follows}{works follow} the simple structure
\begin{equation}
	R_\theta = U \circ FC \circ S_L \circ S_{L-1} \circ \ldots \circ S_2 \circ S_1 \circ \crop,
	\label{eq:reg}
\end{equation}
where \( \Function{\crop}{\R^n}{\R^{\tilde{n}}} \) \deleted{is a} center-crops the validation images of size \( n = \num{640} \times w \) to the size of the training images.
The layer \( S_l \colon x \mapsto \lrelu(\widetilde{W}_l\lrelu(W_l x + b_l) + \tilde{b}_l) \), \( l \in \llbracket 1, L \rrbracket \) utilizes the leaky ReLU \( \lrelu \colon x \mapsto \max\{ \gamma x, x \} \) with leak coefficient \( \gamma = \num{0.05} \).
The weights and biases \( \{ W_l, b_l, \widetilde{W}_l, \tilde{b}_l \} \) in the \( l^{\text{th}} \) layer encode the convolution kernels of size \( \num{3} \times \num{3} \) where \( \widetilde{W}_{l} \) has stride \num{2}.
The number of features in each of the \( L = 6 \) layers follows a geometric progression with common ratio \num{1.75}, starting at \num{48} features in the first layer.
Finally, \( FC \colon x \mapsto W_{\text{FC}}x \) is a fully connected layer mapping to a scalar and \( U \) is the absolute value.
The entirety of the learnable parameters can be summarized as \( \theta = {\{ W_l, b_l, \widetilde{W}_l, \tilde{b}_l \}}_{l=1}^L \cup \{ W_{\text{FC}} \} \).
We do not impose any constraints on any of the parameters, thus \( \Theta \cong \R^{n_p} \), where \( n_p = \num{21350640} \) is the total number of parameters.
\added{%
	Although our network is quite sizable, it has significantly less learnable parameters than, e.g., the discriminative end-to-end \gls{vn} of~\cite{sriram_endtoend_2020} (\num{3e7}) or the score-based diffusion models of~\cite{chung_scoremri_2022} (\num{6.7e7}).%
}

We optimize~\cref{eq:ml} with AdaBelief~\cite{zhuang2020adabelief} (\( \beta_1 = 0.9, \beta_2 = 0.999 \)).
\added{%
	In contrast to most previous works~\cite{du_implicit_2019,Guan_energy_2022,tu_collaborative_2023} we did not find it necessary to regularize our model by means of, e.g., Lipschitz regularization, magnitude penalization or similar techniques.
}
We use a learning rate of \( \num{5e-4} \), exponentially decreasing with rate \num{0.5} at update steps \( \{ \num{500}, \num{2000}, \num{3000}, \num{5000}, \num{7000} \} \), using a batch size of \( 50 \) for \num{27000} parameter updates.
To stabilize training, we smooth the data distribution by convolving it with a normal distribution of standard deviation \( \num{1.5e-2} \).
For approximating \( \mathbb{E}_{p_\theta} \), we run \gls{ula} for \( J_{\text{max}} = 500 \) steps.
To accelerate training in the early stages, we use an exponential schedule, detailed by \( J_h = \lceil J_{\text{max}} (1 - \exp(\sfrac{-h}{\num{1000}})) \rceil \), at the \( h^{\text{th}} \) parameter update.
\replaced{%
	For persistent initialization, we use a replay buffer holding \num{8000} images with reinitialization chance \( \pi_{\text{reinit}} = \qty{1}{\percent} \).
}{
	Further, to minimize the burn-in time of our sampler, we use a replay buffer~\cite{tieleman_persistent_2008} holding 8000 images.
	The reinitialization chance of the replay buffer is 1\%, and reinitialization happens with an equal chance of uniform noise or samples from the dataset.
}
Training took approximately one month on a machine equipped with one NVIDIA Quadro RTX 8000.
\subsection{Simulation study and posterior sampling}%
\label{ssec:simulation study}
To construct a real-valued simulation study, we use the \gls{rss} validation and test data detailed in~\cref{ssec:experimental data}, and retrospectively sub-sample the corresponding \( k \)-space data.
To approximately map the reconstructions to the same intensities seen during training, we normalize the data by \( z \mapsto \sfrac{z}{\norm{u^{(0)}}_\infty} \), and normalize the final reconstruction by \( u^{*} \mapsto u^{*}\norm{u^{(0)}}_\infty \).
To run inference on this data, we utilize~\cref{alg:ipalm} with one sensitivity map that is fixed at the identity (\( C = 1 \), \( \sigma_1 = \mathbf{1} \) is fixed as the one-vector).
We found the optimal regularization parameter \( \lambda \) by grid search on the validation dataset.

In addition to computing \gls{map} estimators in the sense of~\cref{eq:mri problem}, we also examine the posterior distribution.
In particular, we approximate the \gls{mmse} (\cref{eq:mmse}) and variance (\cref{eq:variance}) integrals by \gls{mcmc}:
We run \gls{ula} on the Gibbs density of \( H \), where again \( C = 1 \) and \( \sigma_1 = \mathbf{1} \) is fixed as the one-vector, i.e.\ \( p(u \mid z) \propto \exp\big( -H(u, \mathbf{1}) \big) \).
The algorithm is initialized with uniform noise.
We discard the first \num{10000} samples to reach a steady state, and then save every \replaced{\nth{15}}{15} iteration, for a total of \num{160000} iterations (resulting in \num{10000} saved samples).
We found that the regularization parameter \( \lambda \) barely influenced the results of the posterior sampling and consequently set it to \( \lambda = 1 \) for all sub-sampling \replaced{patterns}{masks}.
\subsection{Parallel imaging}\label{ssec:methods pi}
For the parallel imaging experiments, we run~\cref{alg:ipalm} with \( K = 100 \).
As in the real-valued simulation study, we normalize the data by \( z \mapsto \sfrac{z}{\norm{u^{(0)}}_\infty} \), and normalize the final reconstruction by \( u^{*} \mapsto u^{*}\norm{u^{(0)}}_\infty \).
To find the optimal regularization parameters, we fix \( \mu = 10 \) and obtain \( \lambda \) by linear least-squares regression of the initial residuum \( \sum_{c=1}^C \norm{\mathcal{F}_M(\sigma_{c}^{(0)}(z^{\text{val},i}) \odot u^{(0)}(z^{\text{val},i})) - {(z^{\text{val},i})}_{c}}_2^2 \) against \( \min_{\lambda} \norm{\optimal{u}(z^{\text{val},i}, \lambda) - u^{\text{val},i}}_2^2 \) (found by grid search) for all image-data pairs \( (u^{\text{val},i}, z^{\text{val},i}) \) in the validation set.
Here, we view the initial image \( u^{(0)} \), the sensitivity maps \( {(\sigma_c)}_{c=1}^C \), and the optimal reconstruction \( u^* \) as maps to make dependencies explicit.
This regression is performed \emph{only} for \num{4}-fold Cartesian sub-sampling with \qty{8}{\percent} \gls{acl} and other sub-sampling \replaced{patterns}{schemes} use the same fit.
Generalization experiments marked with \( \dagger \) use the linear \( \lambda \)-fit calculated on \gls{corpd} data.
For experiments marked with \( * \), we re-ran the regression on \gls{corpdfs} data on a \num{4}-fold Cartesian sub-sampling with \qty{8}{\percent} \gls{acl}.

A particular characteristic of our reconstruction approach is that \replaced{its}{it} intensities are not quantitatively comparable to the reference.
In detail, although we normalize the reconstruction by the \gls{rss} of the sensitivity maps, we observed that especially in low-intensity regions (e.g.\ air) the reconstruction did not match the reference.
To remedy this and allow for fair quantitative evaluation, we utilize the validation data to fit a spline curve (cubic splines, \num{5} equally spaced knots) against the scatter of reconstructed and reference intensities.
For the generalization experiments, we fit the spline curve again on an independent \gls{corpdfs} validation dataset.
\added{%
	The spline curves for both CORPD and CORPD-FS are shown in~\cref{fig:splines} in the appendix.
	The insets show that our reconstructions prefer zero-intensity in background regions, whereas the reference images have non-zero background intensity.
}

For parallel imaging, we define the \gls{mmse} as follows:
Let \( (\optimal{u}, {(\optimal{\sigma}_c)}_{c=1}^C) \) be a solution to~\cref{eq:mri problem}.
Then, we fix \( {(\optimal{\sigma}_c)}_{c=1}^C \) and sample the conditional probability \( p(u \mid z, {(\optimal{\sigma}_c)}_{c=1}^C ) \).
This amounts to performing Langevin sampling on the Gibbs density of \( H(\, \cdot \,, {(\optimal{\sigma}_c)}_{c=1}^C) \colon \R_{\geq0}^n \to \R \).
\added{%
	The sensitivities may also be included in the Langevin procedure, but we empirically found no noticeable difference to freezing them.
	We believe that this is due to the strong imposed spatial regularity.
}

\added{%
	We evaluate the quality of the estimated sensitivity maps by computing the null-space residual~\cite{uecker_espirit_13}:
	Let \( u_c = \mathcal{F}^\ast(z_c) \), \( c = 1,\dotsc,C \), denote the fully-sampled coil images.
	The null-space residual
	\begin{equation}
		\pi_c = \frac{\sigma_c}{|\Sigma|^2_{\mathcal{C}}} \sum_{i=1}^{C} \bar{\sigma}_i u_i - u_c,
		\label{eqq:null space residual}
	\end{equation}
	where \( \bar{\,\cdot\,} \) denotes complex conjugation, should only contain noise since \( u_i = \sigma_i u \) when \( \sigma_i \) is exact.
	Thus, any residual signal points to sub-optimal sensitivity estimates.
}
\subsection{Comparison and evaluation}
We compare our approach to the following methods:
For a hand-crafted prior, we chose the Charbonnier smoothed \replaced{total variation
\begin{equation}
	\text{\gls{tv}}: \R^n_{\geq 0} \ni x \mapsto \sum_{i=1}^n \sqrt{{(\mathrm{D} x)}_{i}^2 + {(\mathrm{D} x)}_{i + n}^2 + \epsilon^2}
	\label{eq:charb}
\end{equation}}{\gls{tv}: \( \R^n_{\geq 0} \ni x \mapsto \sum_{i=1}^n \sqrt{{(\mathrm{D} x)}_{i}^2 + {(\mathrm{D} x)}_{i + n}^2 + \epsilon^2} \),}
with \( \epsilon = \num{e-3} \).
In the real-valued simulation study, we compare against the fastMRI baseline method~\cite{zbontar_fastmri_2018}
\added{%
	as well as the diffusion-based approach of~\cite{chung_scoremri_2022}.
	However, due to time and computational constraints, we limit our comparison to one arbitrarily picked image per sub-sampling pattern.
	The implementation as well as the trained model are taken from their github repository, thus the training database differs to ours as it includes the CORPD-FS data.
	We use \num{2000} steps in the reverse diffusion.
}
As a state-of-the-art discriminative approach for parallel \gls{mri}, we compare against the end-to-end \gls{vn} approach from \cite{sriram_endtoend_2020}.
The implementation was taken from the fastMRI github repository with default parameters.
The \replaced{fastMRI baseline method as well as the \gls{vn}}{models} were trained on the subset of the fastMRI dataset detailed~\cref{ssec:experimental data}, with masks generated using random \num{4}-fold Cartesian sub-sampling with \qty{8}{\percent} \gls{acl}.

We compare the reconstructions quantitatively using the \gls{psnr}, \gls{nmse} and \gls{ssim}~\cite{zhou_ssim_2004}.
\gls{ssim} uses a \( \num{7} \times \num{7} \) uniform filter and parameters \( K_1 = \num{0.01} \), \( K_2 = \num{0.03} \).
We define the acceleration factor (denoted Acc.\ in tables showing quantitative results) as the ratio of the image size and the acquired \( k \)-space points on the Cartesian grid.
\section{Results}
\subsection{Data-independent analysis}
The local regularization landscape is a proxy for many interesting properties, relating to (local) convexity, generalization capabilities and adversarial robustness~\cite{Stutz_2021_adversarial}.
To effectively visualize the \( \tilde{n} = (\num{320} \times \num{320}) \)-dimensional landscape, we follow~\cite{li_2018_visualizing}:
Let \( {(x^{(k)})}_{k=1}^{K} \), \( x^{(k)} \in \R^{\tilde{n}} \) be samples drawn from the Langevin process~\cref{eq:langevin}, starting from uniform noise.
Denote with \( (v_1, v_2) \) the first two principal components of the matrix \( [x^{(0)} - x^{(K)};x^{(1)} - x^{(K)};\dotsc;x^{(K-1)} - x^{(K)} ] \) and denote \( \bar{x} \coloneqq x^{(K)} + \sfrac{1}{K-1} \sum_{k=1}^{K-1} (x^{(k)} - x^{(K)}) \).
For the experiments, we set \( K = \num{10000} \).

\begin{figure*}
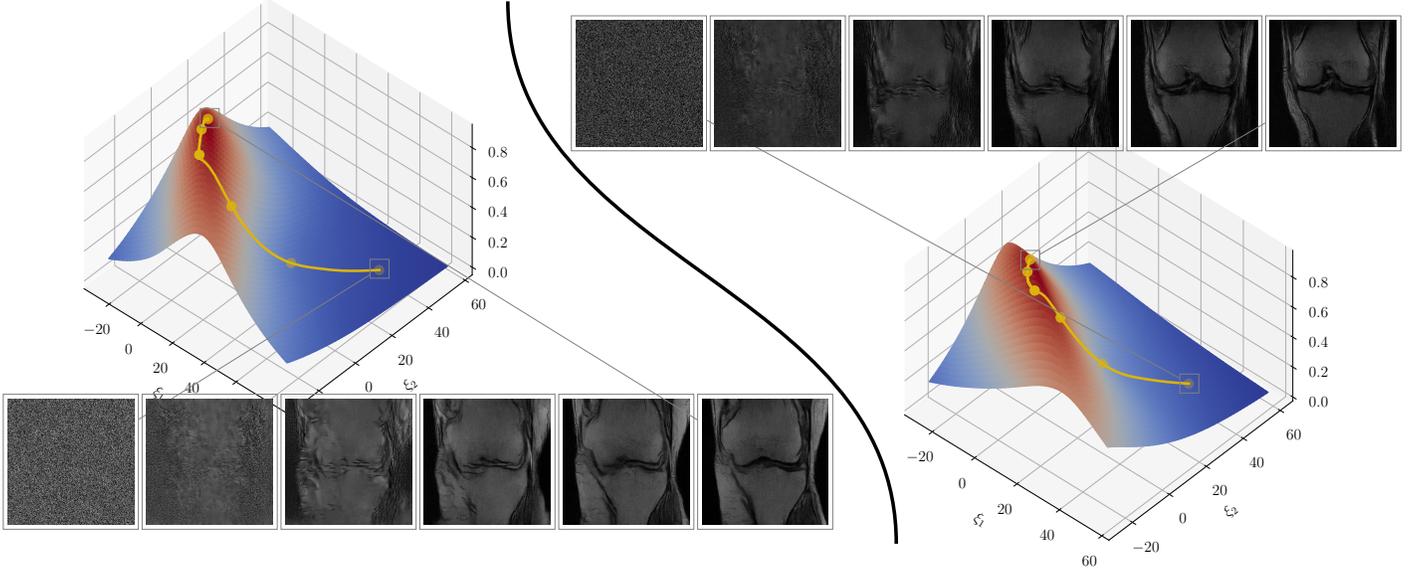

	\def\prefix{./figures/trajectories/}
	\centering
	\resizebox{\textwidth}{!}{%
	\begin{tikzpicture}
		\foreach [count=\icherry] \cherry in {0, 1}
		{
			\node at (\icherry * 13-5, -\icherry * 2) {\includegraphics[trim={2.1cm, 1cm, 1.0cm, 1cm}, clip, width=7cm]{\prefix/\cherry/pdf.pdf}};
			\foreach [count=\iindex] \ind/\coords in {%
				0/{{"9.49, -2.94", "22.32, -4.76"}},%
				20/{{"10.6, -.86", "19.6, -.84"}},%
				50/{{"10.6, -.86", "19.6, -.84"}},%
				200/{{"10.6, -.86", "19.6, -.84"}},%
				500/{{"7.85, 0.98", "16.85, 0.93"}},%
				999/{{"6.8, -.55", "19.8, -2.80"}}%
			}
			{
				\pgfmathsetmacro{\yoff}{((\icherry - 2) * 2 + 1) * 3 - 3}
				\node (sample\icherry\iindex) [draw, rectangle, inner sep=2, gray] at (\icherry * 9 + \iindex * 2.2 - 3 * 2.2, \yoff) {\includegraphics[angle=180,origin=c,width=2cm]{\prefix/\cherry/\ind.png}};
				\pgfmathparse{\coords[\icherry-1]}
				\ifthenelse{\iindex=1 \OR \iindex=6}{
					\node (anno\icherry\iindex) [draw, gray, rectangle, minimum width=0.3cm, minimum height=0.3cm] at (\pgfmathresult) {};
					\draw [gray] (sample\icherry\iindex) -- (anno\icherry\iindex);
				}{}
			}
		}
		\draw [ultra thick] ([xshift=-1cm,yshift=1.3cm]sample21.west) to [out=270,in=90] ([xshift=1cm,yshift=-1.3cm]sample16.east);
	\end{tikzpicture}%
	}
	\caption{%
		Surface plots of the local regularization landscape \( \R^2 \ni (\xi_1, \xi_2) \mapsto \exp(-R(\bar{x} + \xi_1v_1 + \xi_2v_2)/\widetilde{T}) \).
		The Langevin trajectory (\cref{eq:langevin}) is shown in gold along with some representative states.
	}%
	\label{fig:pdf sampling}
\end{figure*}

In~\cref{fig:pdf sampling} we show \( \R^2 \ni (\xi_1, \xi_2) \mapsto \exp \big( -R(\bar{x} + \xi_1 v_1 + \xi_2 v_2) / \widetilde{T} \big) \), where \( \widetilde{T} = 7 \) was chosen to yield visually pleasing results, along with the Langevin trajectory.
The figure shows two interesting properties:
First, the samples from the Langevin process are almost indistinguishable from the reference data, which demonstrates that our prior is extremely strong and faithfully represents the distribution of the training data.
\added{%
	In particular, this implies that the learned regularizer --- by construction --- should only be applied to reconstruction tasks where the underlying data is also drawn from the same distribution.
	The experiments in~\cref{ssec:generalization} show that our regularizer can reasonably reconstruct knee images of different contrast by adapting regularization strength, but prior work~\cite{zach_generative_2021} demonstrates that performance quickly degrades when confronted with, e.g., rotated images.
}
Second, in two dimensions, the landscape appears smooth on the considered domain and almost log-concave around modes of the induced distribution.
Although not applicable directly, these findings should be taken as evidence that the high-dimensional landscape is also reasonably well-behaved.
This is corroborated empirically by the ease of optimization: For all reconstruction tasks, we only need in the order of \num{10} iterations of \gls{ipalm}.
\subsection{Simulation Study}
\begin{figure*}
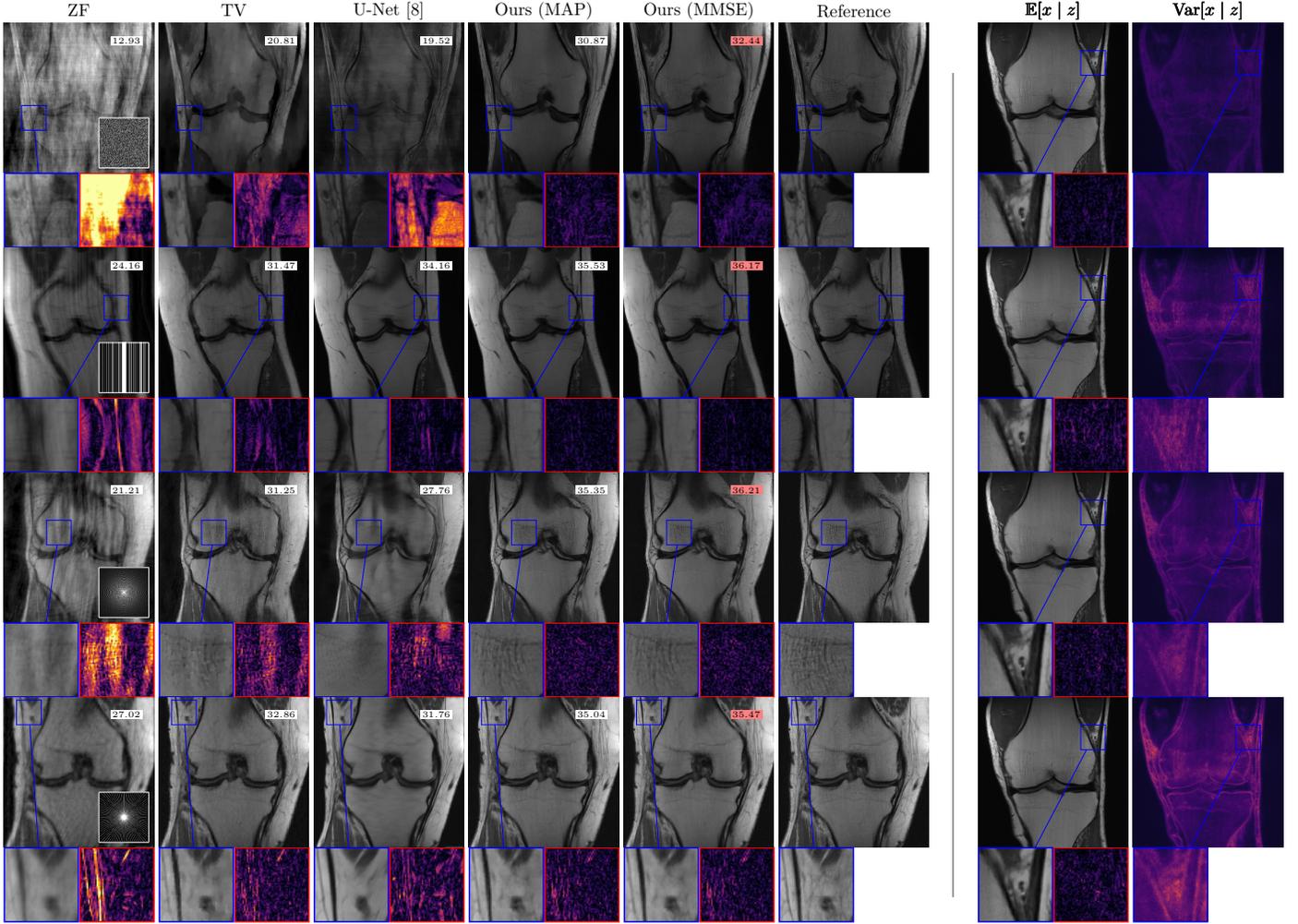

	\def\prefix{./figures/single-coil-synthetic}
	\def\cherries{{"file1002546.h5","file1001331.h5","file1001687.h5","file1002351.h5"}}
	\def\spyoff{{{-0.9,-0.4},{0.75,0.3},{-0.4,0.3},{-1,1.2}}}
	\def\TableQuant{{
		{12.931758880615234, 20.806852340698242, 19.52480125427246, 30.8723201751709, 32.44416046142578},
		{24.160869598388672, 31.469541549682617, 34.161155700683594, 35.53403091430664, 36.168399810791016},
		{21.210956573486328, 31.248300552368164, 27.76167869567871, 35.35053253173828, 36.209625244140625},
		{27.015140533447266, 32.86052322387695, 31.76373863220215, 35.04103088378906, 35.46697998046875}
	}}
	\def\BestTag{{5, 5, 5, 5}}
	\centering
	\resizebox{\textwidth}{!}{%
	\begin{tikzpicture}
		\foreach [count=\isampling] \sampling in {random, cartesian, spiral, radial}
		{
			\pgfmathsetmacro{\cherry}{\cherries[\isampling-1]}
			\foreach [count=\imethod] \method/\manno in {{zero_filled}/\gls{zf}, tv/TV, unet/{U-Net~\cite{zbontar_fastmri_2018}}, ours/{Ours (\gls{map})}, {mean}/{Ours (\gls{mmse})}, ground_truth/Reference}
			{
				\ifthenelse{\isampling=1}{
					\node at (3.1*\imethod, -2.75) {\manno};
				}{}
				\ifthenelse{\imethod=5}{%
					\def\ourpath{\prefix/posterior/\sampling/\cherry}
				}{%
					\def\ourpath{\prefix/\sampling/\cherry}
				}
				\pgfmathsetmacro{\spyxoff}{\spyoff[\isampling-1][0]}
				\pgfmathsetmacro{\spyyoff}{\spyoff[\isampling-1][1]}
				\coordinate (onn) at (\imethod * 3.1 + \spyxoff, -\isampling*4.5 + \spyyoff);
				\ifthenelse{\imethod=6}{}{
				\begin{scope}[spy using outlines={rectangle, magnification=3, width=1.48cm, height=1.48cm, connect spies}]
					\node at (3.1*\imethod, -\isampling*4.5) {\includegraphics[angle=180,origin=c,width=3cm]{\ourpath/d_\method.png}};
					\spy [red] on (onn) in node [left] at (\imethod * 3.1 + 1.5, -\isampling*4.5-2.25);
				\end{scope}}
				\begin{scope}[spy using outlines={rectangle, magnification=3, width=1.48cm, height=1.48cm, connect spies}]
					\node at (3.1*\imethod, -\isampling*4.5) {\includegraphics[angle=180,origin=c,width=3cm]{\ourpath/\method.png}};
					\spy [blue] on (onn) in node [left] at (\imethod * 3.1-0.01, -\isampling*4.5-2.25);
				\end{scope}
				\ifthenelse{\imethod<6}{
					\pgfmathsetmacro{\psnr}{\TableQuant[\isampling-1][\imethod-1]}
					\pgfmathsetmacro{\ibest}{\BestTag[\isampling-1]}
					\ifthenelse{\imethod=\ibest}{\def\ccolor{white!50!red}}{\def\ccolor{white}}
					\node [inner sep=0.1em, outer sep=0.1em, fill=\ccolor] at (3.1*\imethod+0.95, -\isampling*4.5+1.15) {\tiny\num[round-mode=places,round-precision=2]{\psnr}};
				}{}
			}
			\node at (4, -\isampling*4.5-0.9) {\includegraphics[cframe=white,height=1cm]{\prefix/\sampling/mask.png}};
		}
	\end{tikzpicture}
	\def\prefix{./figures/posterior}
	\def\cherry{file1002382.h5}
	\begin{tikzpicture}
		\draw [thick, black!50] (1.1, -4) -- ++(0, -16.5);
		\foreach [count=\isampling] \sampling in {random, cartesian, spiral, radial}
		{
			\foreach [count=\iimname] \imname/\manno in {mean/{\(\mathbb{E}[x\mid z]\)}, var/{\( \operatorname{Var}[x\mid z] \)}} {
				\node at (3.1*\iimname, -2.75) {\manno};
				\ifthenelse{\iimname=1}{
				\begin{scope}[spy using outlines={rectangle, magnification=3, width=1.48cm, height=1.48cm, connect spies}]
					\node at (3.1*\iimname, -\isampling*4.5) {\includegraphics[angle=180,origin=c,width=3cm]{\prefix/\sampling/\cherry/error.png}};
					\spy [red] on (\iimname * 3.1 + 0.8, -\isampling*4.5 + 0.7) in node [left] at (\iimname * 3.1 + 1.5, -\isampling*4.5-2.25);
				\end{scope}}{}
				\begin{scope}[spy using outlines={rectangle, magnification=3, width=1.48cm, height=1.48cm, connect spies}]
					\node at (3.1*\iimname, -\isampling*4.5) {\includegraphics[angle=180,origin=c,width=3cm]{\prefix/\sampling/\cherry/\imname.png}};
					\spy [blue] on (\iimname * 3.1 + 0.8, -\isampling*4.5 + 0.7) in node [left] at (\iimname * 3.1 - 0.01, -\isampling*4.5-2.25);
				\end{scope}
			}
		}
	\end{tikzpicture}%
	}
	\caption{%
		Real-valued simulation study results:%
			\nth{1} row:
			Random sub-sampling with acceleration factor \num{3},
			\nth{2} row:
			\num{4}-fold Cartesian sub-sampling with \qty{8}{\percent} \gls{acl},
			\nth{3} row:
			Spiral sub-sampling (\num{5}-fold acceleration),
			\nth{4} row:
			Radial sub-sampling using \num{45} spokes.
		The inlays show the sub-sampling \replaced{pattern}{mask} (white), a detail zoom (blue), and the magnitude of the difference to the reference (red, \num{0}~\protect\drawcolorbar~\num{0.2}).
		Left: Reconstruction results.
		The numbers show \gls{psnr} (over the entire test dataset), with the best method emphasized in red.
		Right: Uncertainty quantification through marginal posterior variance (\( \num{0}~\protect\drawcolorbar~\num{0.0025} \)).
	}%
	\label{fig:simulation study}
\end{figure*}

We show qualitative results of the retrospectively sub-sampled real-valued data in~\cref{fig:simulation study} (left), where the results are shown in increasing acceleration.
In the second row we show the results from \num{4}-fold Cartesian sub-sampling with \qty{8}{\percent} \gls{acl}.
Note that this is the setting on which the U-Net was trained, and indeed it yields satisfactory reconstructions.
The \gls{tv} reconstruction for this task is not able to fully remove the sub-sampling artifacts, but increasing regularization would lead to significant loss of detail.
Our approach is at least on par qualitatively with the discriminative U-Net, and the quantitative analysis in~\cref{tab:simulation results} shows superiority over all reference methods.

The third row details the reconstructions for spiral sub-sampling.
Here, the acceleration factor is approximately \( \num{5} \), with a more densely sampled \( k \)-space center, which is typically advantageous for traditional reconstruction techniques.
The \gls{tv} reconstruction removes most of the sub-sampling artifacts, although some are still visible, especially in the background.
The discriminative U-Net approach in this task (and indeed all tasks apart from Cartesian sub-sampling, on which it was trained) struggles to discriminate between details in the anatomy and sub-sampling artifacts.
Thus, it hallucinates details into the reconstruction that are not reflected in the data.
In contrast, our approach is able to faithfully reconstruct the knee with no visible artifacts.
The results are similar for the pseudo-radial sub-sampling pattern with \num{45} spokes shown in the fourth row.

The \num{3}-fold random sub-sampling shown in the first row is particularly interesting.
Here, the \( k \)-space center is not more densely sampled than any other region, which manifests in the zero-filling reconstruction by large-scale intensity shifts.
None of the reference methods are able to correct this, since they do not have knowledge of the anatomy of the human knee.
On the other hand, our approach, due to its generative nature, can restore the general shape of the knee well, and due to the variational approach also faithfully keeps details that are present in the data.
The generative approach alleviates the requirement for sampling schemes to densely sample the \( k \)-space center, which is in stark contrast to the general theory.%

\added{%
	We show a quantitative comparison against the diffusion-based approach of~\cite{chung_scoremri_2022} in~\cref{tab:diffusion comparison} that includes reconstruction time and number of trainable parameters.
	The accompanying qualitative results are shown in~\cref{fig:diffusion comparison} in the appendix.
	The results are separated since we only evaluated on one image per sub-sampling pattern due to time and computational constraints, and thus it does not constitute a comprehensive comparison.
	Our \gls{mmse} estimate consistently beats the diffusion-based approach, and our \gls{map} estimate is only inferior for the random sub-sampling pattern.
	To generate the \num{10000} samples for the \gls{mmse} estimate takes only about \qty{30}{\percent} more time compared to the one sample generated by the reverse \gls{sde}.
	On the other hand, computing our \gls{map} estimate is about \num{80} times faster.
}
\begin{table}
\centering
\begin{threeparttable}
	\caption{%
		Quantitative results for the real-valued simulation study using different \( k \)-space trajectories.
		Numbers in parenthesis indicate the acceleration factor, bold typeface indicates the best method.
	}%
	\label{tab:simulation results}
	\setlength{\tabcolsep}{0.5em}
	\begin{tabular}{ll*{5}{S[text-series-to-math,table-format=2.2,round-mode=places,round-precision=2]}}
		\toprule
		& & {\multirow{2}{*}{\gls{zf}}} & {\multirow{2}{*}{\gls{tv}}} & {\multirow{2}{*}{U-Net}} & \multicolumn{2}{c}{Ours} \\
		\cmidrule(l{1em}r{1em}){6-7}
		& & & & & {\small\gls{map}} & {\small\gls{mmse}} \\
		\toprule
		\multirow{3}{*}{Random (\num{3})} %
		& \( \mathparagraph \) & 12.931758880615234 & 20.806852340698242 & 19.52480125427246 & 30.8723201751709 & \bfseries 32.44416046142578 \\
		& \( \ddagger \)       & \fpeval{100 * 0.6423648595809937} & \fpeval{100 * 0.10220915079116821} & \fpeval{100 * 0.12830260396003723} & \fpeval{100 * 0.013434510678052902} & \bfseries \fpeval{100 * 0.010225472040474415} \\
		& \( \mathsection \)   & 0.4817352294921875 & 0.7321547269821167 & 0.5721480846405029 & 0.8576399087905884 & \bfseries 0.9027918577194214 \\
		\midrule
		\multirow{3}{*}{Cartesian (\num{4})} %
		& \( \mathparagraph \) & 24.160869598388672 & 31.469541549682617 & 34.161155700683594 & 35.53403091430664 & \bfseries 36.168399810791016 \\
		& \( \ddagger \)       & \fpeval{100 * 0.059642061591148376} & \fpeval{100 * 0.008760045282542706} & \fpeval{100 * 0.004405752755701542} & \fpeval{100 * 0.003225221298635006} & \bfseries \fpeval{100 * 0.002839048160240054} \\
		& \( \mathsection \)   & 0.704010546207428 & 0.853325366973877 & 0.8865730166435242 & 0.8949872851371765 & \bfseries 0.9105123281478882 \\
		\midrule
		\multirow{3}{*}{Spiral (\( \approx \num{5} \))} %
		& \( \mathparagraph \) & 21.210956573486328 & 31.248300552368164 & 27.76167869567871 & 35.35053253173828 & \bfseries 36.209625244140625 \\
		& \( \ddagger \)       & \fpeval{100 * 0.1024087592959404} & \fpeval{100 * 0.009196978993713856} & \fpeval{100 * 0.019011199474334717} & \fpeval{100 * 0.0033660277258604765} & \bfseries \fpeval{100 * 0.0028191949240863323} \\
		& \( \mathsection \)   & 0.617068350315094 & 0.84514981508255 & 0.7843644618988037 & 0.8829594850540161 & \bfseries 0.9014105200767517 \\
		\midrule
		\multirow{3}{*}{Radial (\( \approx \num{6} \))} %
		& \( \mathparagraph \) & 27.015140533447266 & 32.86052322387695 & 31.76373863220215 & 35.04103088378906 & \bfseries 35.46697998046875 \\
		& \( \ddagger \)       & \fpeval{100 * 0.03293876722455025} & \fpeval{100 * 0.006426490843296051} & \fpeval{100 * 0.007553369738161564} & \fpeval{100 * 0.0036013659555464983} & \bfseries \fpeval{100 * 0.0033337108325213194} \\
		& \( \mathsection \)   & 0.4817352294921875 & 0.7103990912437439 & 0.6549839973449707 & 0.8577146530151367 & \bfseries 0.8871539831161499 \\
		\bottomrule
	\end{tabular}
	\begin{tablenotes}\footnotesize
		\( \mathparagraph \): \gls{psnr} (\si{\decibel}) \( \uparrow \), \( \ddagger \): \gls{nmse} (\( \times \num{e2} \)) \( \downarrow \), \( \mathsection \): \gls{ssim} \( \uparrow \)
	\end{tablenotes}
\end{threeparttable}
\end{table}%
\begin{table}
	\caption{%
		\added{%
			Quantitative comparison (\gls{psnr}) against the diffusion-based method of~\cite{chung_scoremri_2022}.
			Bold typeface indicates the best method.
		}
	}%
	\label{tab:diffusion comparison}
	\resizebox{\columnwidth}{!}{
	\begin{tabular}{llcccccc}
		& & Random & Cartesian & Spiral & Radial & Time (s) & Parameters \\\toprule
		\multirow{2}{*}{Ours} & \gls{map} & 31.78 & 34.97 & 36.07 & 36.02 & 3.13 & \multirow{2}{*}{\num{2.1e7}}\\\cmidrule{2-7}
							  & \gls{mmse} & \textbf{34.76} & \textbf{35.55} & \textbf{37.00} & \textbf{36.56} & 333.3 &\\\midrule
		\cite{chung_scoremri_2022} & & 33.31 & 34.65 & 35.66 & 34.98 & 251.6 & \num{6.8e7} \\\bottomrule
\end{tabular}}
\end{table}
\subsection{Uncertainty Quantification through Posterior Sampling}
The natural probabilistic interpretation of our variational approach gives rise to a distribution of reconstructions for any particular reconstruction problem.
We exploit this distribution to compute pixel-wise variance maps as well as the \gls{mmse} estimate.
The results in~\cref{fig:simulation study} (right) show large variance around small structures, and more variance when less data is available:
The three-fold random sub-sampling shows the least variance, while the approximately six-fold radial sub-sampling shows the most.
The \gls{mmse} estimator also yields visually pleasing reconstructions.
In fact, \cref{tab:simulation results} \added{and~\cref{tab:diffusion comparison}} establish\deleted{es} quantitative superiority over the \gls{map} estimate (see~\cref{sec:discussion} for a possible explanation).
\subsection{Parallel Imaging}%
\begin{figure*}
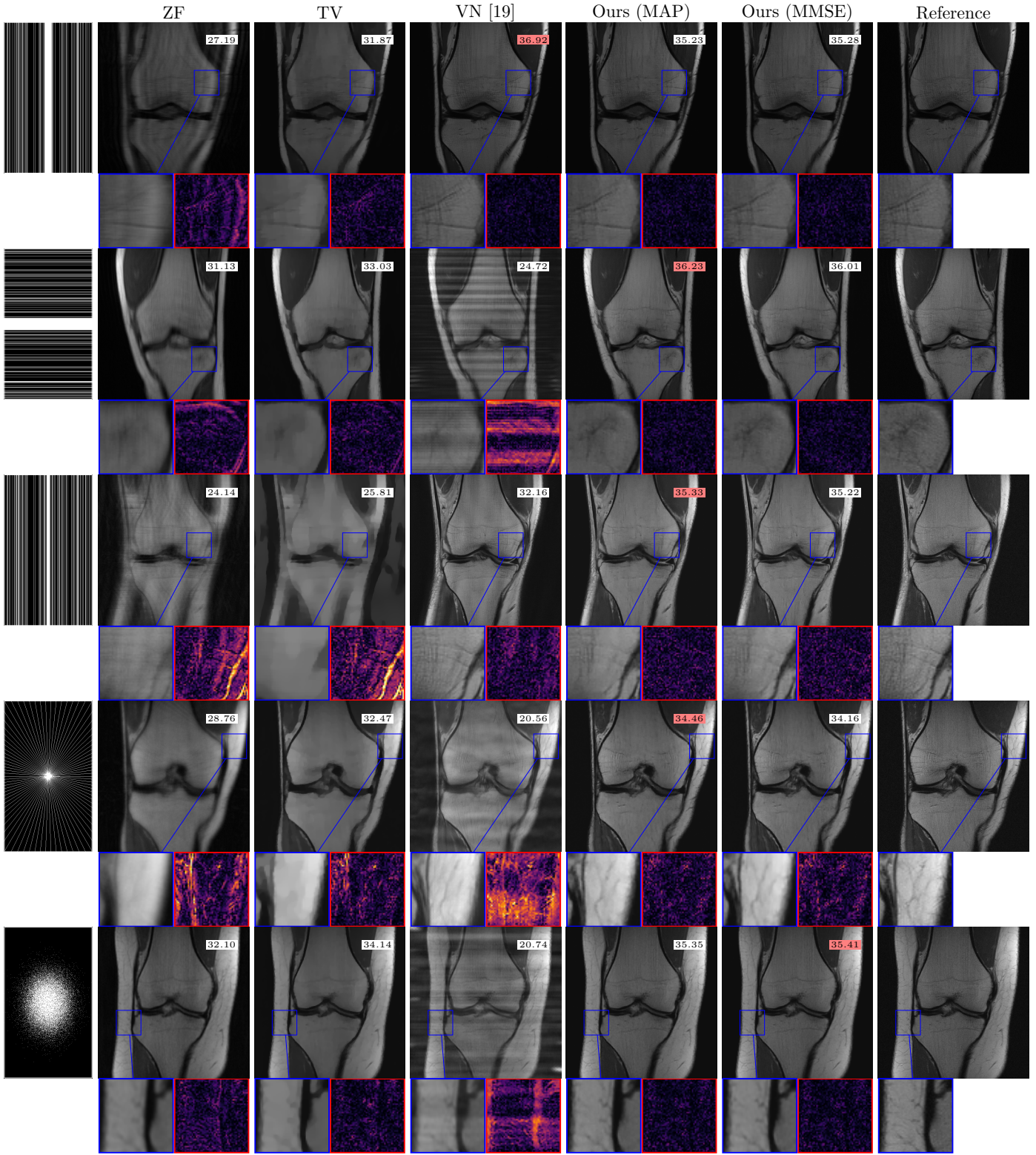

	\def\prefix{./figures/validation}
	\def\cherries{{"file1001057.h5","file1002021.h5","file1001598.h5","file1001983.h5","file1002257.h5"}}
	\def\spyoff{{{0.65,0.3},{0.6,-0.7},{0.5,0.1},{1.2,0.6},{-0.9,-0.4}}}
	\def\TableQuant{{
		{27.18990707397461, 31.867795944213867, 36.92045211791992, 35.22513961791992, 35.27544403076172},
		{31.12578582763672, 33.03041076660156, 24.720752716064453, 36.230690002441406, 36.01150894165039},
		{24.141496658325195, 25.81011390686035, 32.16292190551758, 35.33468246459961, 35.224971771240234},
		{28.76137351989746, 32.46963882446289, 20.558855056762695, 34.45677947998047, 34.1624641418457},
		{32.0957145690918, 34.1363525390625, 20.741687774658203, 35.34770965576172, 35.412174224853516}
	}}
	\def\BestTag{{3, 4, 4, 4, 5}}
	\centering
	\resizebox{\textwidth}{!}{%
	\begin{tikzpicture}
		\foreach [count=\isampling] \sampling in {cartesian, cartesian_rot90, {cartesian_4}, radial, gaussian2d}
		{
			\pgfmathsetmacro{\cherry}{\cherries[\isampling-1]}
			\foreach [count=\imethod] \method/\manno in {{zero_filled}/\gls{zf}, tv/TV, vn/{\gls{vn}~\cite{sriram_endtoend_2020}}, ours/{Ours (\gls{map})}, ours_mmse/{Ours (\gls{mmse})},ground_truth/Reference}
			{
				\ifthenelse{\isampling=1}{
					\node at (3.1*\imethod, -2.8) {\manno};
				}{}
				\pgfmathsetmacro{\spyxoff}{\spyoff[\isampling-1][0]}
				\pgfmathsetmacro{\spyyoff}{\spyoff[\isampling-1][1]}
				\coordinate (onn) at (\imethod * 3.1 + \spyxoff, -\isampling*4.5 + \spyyoff);
				\ifthenelse{\imethod=6}{}{
				\begin{scope}[spy using outlines={rectangle, magnification=3, width=1.47cm, height=1.47cm, connect spies}]
					\node at (3.1*\imethod, -\isampling*4.5) {\includegraphics[angle=180,origin=c,width=3cm]{\prefix/\sampling/\cherry/d_\method.png}};
					\spy [red] on (onn) in node [left] at (\imethod * 3.1 + 1.5 - 0.015, -\isampling*4.5-2.25);
				\end{scope}}
				\begin{scope}[spy using outlines={rectangle, magnification=3, width=1.47cm, height=1.47cm, connect spies}]
					\node at (3.1*\imethod, -\isampling*4.5) {\includegraphics[angle=180,origin=c,width=3cm]{\prefix/\sampling/\cherry/\method.png}};
					\spy [blue] on (onn) in node [left] at (\imethod * 3.1 - 0.015, -\isampling*4.5-2.25);
				\end{scope}
				\ifthenelse{\imethod<6}{
					\pgfmathsetmacro{\psnr}{\TableQuant[\isampling-1][\imethod-1]}
					\pgfmathsetmacro{\ibest}{\BestTag[\isampling-1]}
					\ifthenelse{\imethod=\ibest}{\def\ccolor{white!50!red}}{\def\ccolor{white}}
					\node [inner sep=0.1em, outer sep=0.1em, fill=\ccolor] at (3.1*\imethod+0.95, -\isampling*4.5+1.15) {\tiny\num[round-mode=places,round-precision=2]{\psnr}};
				}{}
			}
			\node at (0.6, -\isampling*4.5) {\includegraphics[cframe=white,height=3cm]{\prefix/\sampling/mask.png}};
		}
	\end{tikzpicture}}
	\caption{%
		Parallel imaging on in-distribution data:
		\nth{1} row:
		\num{4}-fold Cartesian sub-sampling with \qty{8}{\percent} \gls{acl},
		\nth{2} row:
		as previous with swapped phase encoding direction,
		\nth{3} row:
		\num{4}-fold Cartesian sub-sampling with \qty{4}{\percent} \gls{acl},
		\nth{4} row:
		Pseudo-radial sampling with \num{45} equidistant spokes (\( \approx \num{11} \)-fold sub-sampling).
		\nth{5} row:
		2D Gaussian sampling (\num{8}-fold sub-sampling).
		The inlays show the sub-sampling mask, a detail zoom (blue), and the magnitude of the difference to the reference (red, \num{0}~\protect\drawcolorbar~\num{0.2}).
	}%
	\label{fig:parallel imaging id}
\end{figure*}

\label{ssec:parallel imaging}
The first row of~\cref{fig:parallel imaging id} shows a classical Cartesian sub-sampling \replaced{pattern}{mask}, with an acceleration factor of \num{4} and using \qty{8}{\percent} \gls{acl}.
This coincides with the training setup of the discriminative end-to-end variational network approach of~\cite{sriram_endtoend_2020}.
Consequently, it shows the best performance quantitatively as well as qualitatively.
Our method also yields competitive results, achieving a \gls{psnr} of \qty[round-mode=places,round-precision=2]{35.22513961791992}{\decibel}.
This is in line with our expectations, as in general we can not expect a generative approach to beat a discriminative counterpart.
However, the strength of our approach becomes apparent when we slightly change the \replaced{sub-sampling pattern}{acquisition mask}:
The performance of the end-to-end \gls{vn} deteriorates significantly when the phase-encoding direction is swapped, or when less \gls{acl} are acquired.
We emphasize that between these tasks the number of acquired data points is fixed.
With these minimal changes in the \replaced{sub-sampling pattern}{acquisition protocol}, the method of~\cite{sriram_endtoend_2020} is no longer able to fully remove back-folding artifacts or introduces severe hallucinations, while our method yields comparable results for all three tasks.
This is also reflected in the quantitative evaluation in~\cref{tab:parallel imaging}, where our method beats the \gls{vn} approach by a significant margin.

Shifting towards radial and 2D Gaussian sub-sampling \replaced{patterns}{masks}, we observe that the \gls{vn} approach introduces severe artifacts.
We believe that this is a combination of the sensitivity estimation sub-network failing as well as the image sub-network being confronted with previously unseen sub-sampling artifacts.
Indeed, for this task the more general \gls{tv} approach is superior to the \gls{vn}, although reconstructions appear over-smoothed.
In line with expectations, our approach reconstructs the image satisfactorily, yielding the best performance quantitatively as well as qualitatively.
\begin{table*}
	\centering
	\caption{%
		Quantitative results for parallel imaging with different sub-sampling \replaced{patterns}{masks} on in- and out-of-distribution data.
		The \( \dagger \) column shows results using the \gls{corpd} \( \lambda \)-fit, while the \( * \) column has \gls{corpdfs}-adapted parameters (see~\cref{ssec:methods pi}).
	}%
	\label{tab:parallel imaging}
	\hfill%
	\begin{tabular}{lrcl@{}}
		\arrayrulecolor{white}\toprule
		& & & \\
		\arrayrulecolor{black}\midrule
		& {\multirow{2}{*}{Acc.}} & {\multirow{2}{*}{\gls{acl}}} & \\
		\arrayrulecolor{white}\cmidrule(l{1em}r{1em}){1-1}
		& & & \\
		\arrayrulecolor{black}\toprule
		\multirow{9}{*}{Cartesian} & \multirow{9}{*}{\num{4}} & \multirow{3}{*}{\qty{8}{\percent}} & \( \mathparagraph \) \\
		& & & \( \ddagger \) \\
		& & & \( \mathsection \) \\
		\cmidrule(l{.75em}){3-4}
		& & \multirow{3}{*}{\qty{8}{\percent} (horiz.)} %
		&     \( \mathparagraph \) \\
		& & & \( \ddagger \) \\
		& & & \( \mathsection \) \\
		\cmidrule(l{.75em}){3-4}
		 & & \multirow{3}{*}{\qty{4}{\percent}} %
		&     \( \mathparagraph \) \\
		& & & \( \ddagger \) \\
		& & & \( \mathsection \) \\
		\midrule
		\multirow{3}{*}{Radial} & \multirow{3}{*}{\( \approx \num{11} \)} & \multirow{3}{*}{---} %
		&     \( \mathparagraph \) \\
		& & & \( \ddagger \) \\
		& & & \( \mathsection \) \\
		\midrule
		\multirow{3}{*}{2D Gaussian} & \multirow{3}{*}{8} & \multirow{3}{*}{---} & \( \mathparagraph \) \\
		& & & \( \ddagger \) \\
		& & & \( \mathsection \) \\
		\bottomrule
	\end{tabular}\hfill%
	\begin{tabular}{@{}*{5}{S[text-series-to-math,table-format=2.2,round-mode=places,round-precision=2]}@{}}
		\toprule
		\multicolumn{5}{c}{In-distribution (\gls{corpd})} \\
		\midrule
		{\multirow{2}{*}{\gls{zf}}} & {\multirow{2}{*}{\gls{tv}}} & {\multirow{2}{*}{\gls{vn}}} & \multicolumn{2}{c}{Ours} \\
		  \cmidrule(l{1em}r{1em}){4-5}
		& & & {\small \gls{map}} & {\small \gls{mmse}} \\
		\toprule
		27.18990707397461 & 31.867795944213867 & \bfseries 36.92045211791992 & 35.22513961791992 & 35.27544403076172\\
		\fpeval{100 * 0.0223829485476017} & \fpeval{100 * 0.007890134118497372} & \bfseries \fpeval{100 * 0.002435196889564395} & \fpeval{100 * 0.0036337128840386868} & \fpeval{100 * 0.003618051065132022} \\
		0.7375055551528931 & 0.8079418540000916 & \bfseries 0.9163976311683655 & 0.88604736328125 & 0.8878628015518188\\
		\cmidrule{1-5}
		31.12578582763672 & 33.03041076660156 & 24.720752716064453 & \bfseries 36.230690002441406 & 36.01150894165039\\
		\fpeval{100 * 0.00931438896805048} & \fpeval{100 * 0.005917737260460854} & \fpeval{100 * 0.04011896252632141} & \bfseries \fpeval{100 * 0.0028249204624444246} & \fpeval{100 * 0.0029919317457824945} \\
		0.8110671043395996 & 0.8260417580604553 & 0.670006275177002 & \bfseries 0.898271918296814 & \bfseries 0.8988670110702515\\
		\cmidrule{1-5}
		24.141496658325195 & 25.81011390686035 & 32.16292190551758 & \bfseries 35.33468246459961 & 35.224971771240234\\
		\fpeval{100 * 0.04508989304304123} & \fpeval{100 * 0.034832630306482315} & \fpeval{100 * 0.0069937678053975105} & \bfseries \fpeval{100 * 0.0035170933697372675} & \fpeval{100 * 0.0036184450145810843} \\
		0.6923396587371826 & 0.6951988935470581 & \bfseries 0.8927095532417297 & \bfseries 0.8880932331085205 & \bfseries 0.8893769383430481\\
		\midrule
		28.76137351989746 & 32.46963882446289 & 20.558855056762695 & \bfseries 34.45677947998047 & 34.1624641418457\\
		\fpeval{100 * 0.015698831528425217} & \fpeval{100 * 0.006666331551969051} & \fpeval{100 * 0.10128451883792877} & \bfseries \fpeval{100 * 0.004226132296025753} & \fpeval{100 * 0.0045379577204585075} \\
		0.7465711236000061 & 0.8061649203300476 & 0.6916904449462891 & \bfseries 0.8565337657928467 & \bfseries 0.8555713295936584\\
		\midrule
		32.0957145690918 & 34.1363525390625 & 20.741687774658203 & 35.34770965576172 & \bfseries 35.412174224853516\\
		\fpeval{100 * 0.007424597162753344} & \fpeval{100 * 0.004535387735813856} & \fpeval{100 * 0.09953409433364868} & \bfseries \fpeval{100 * 0.003436274826526642} & \bfseries \fpeval{100 * 0.003413973143324256} \\
		0.8387000560760498 & 0.8521340489387512 & 0.6813638210296631 & 0.8809531927108765 & \bfseries 0.8852419853210449\\
		\bottomrule
	\end{tabular}\hfill%
	\begin{tabular}{@{}*{5}{S[text-series-to-math,table-format=2.2,round-mode=places,round-precision=2]}}
		\toprule
		\multicolumn{5}{c}{Out-of-distribution (\gls{corpdfs})} \\
		\midrule
		{\multirow{2}{*}{\gls{zf}}} & {\multirow{2}{*}{\gls{tv}}} & {\multirow{2}{*}{\gls{vn}}} & \multicolumn{2}{c}{Ours} \\
		  \cmidrule(l{1em}r{1em}){4-5}
		& & & {\( \dagger \)} & {\( * \)} \\
		\toprule
		26.091176986694336 & 31.302940368652344 & 29.998945236206055 & 30.602453231811523 & \bfseries 31.712961196899414 \\
		\fpeval{100 * 0.05352313444018364} & \fpeval{100 * 0.01478771585971117} & \fpeval{100 * 0.023820199072360992} & \fpeval{100 * 0.019470086321234703} & \bfseries  \fpeval{100 * 0.013532079756259918} \\
		0.6760171055793762 & 0.7300366759300232 & \bfseries 0.7710118293762207 & 0.731391966342926 & 0.7327521443367004 \\
		\cmidrule{1-5}
		26.57789421081543 & 31.557456970214844 & 28.57035255432129 & 30.890687942504883 & \bfseries 31.652633666992188 \\
		\fpeval{100 * 0.051486801356077194} & \fpeval{100 * 0.014020383358001709} & \fpeval{100 * 0.028979137539863586} & \fpeval{100 * 0.022375022992491722} & \bfseries \fpeval{100 * 0.013702597469091415} \\
		0.7056003212928772 & 0.7315869927406311 & 0.707637369632721 & \bfseries 0.7516953945159912 & 0.7289502024650574 \\
		\cmidrule{1-5}
		24.9786319732666 & 29.91111183166504 & 29.117982864379883 & 29.933029174804688 & \bfseries 31.255582809448242 \\
		\fpeval{100 * 0.06647732108831406} & \fpeval{100 * 0.020912175998091698} & \fpeval{100 * 0.027873607352375984} & \fpeval{100 * 0.029169296845793724} & \bfseries \fpeval{100 * 0.015278858132660389} \\
		0.6545588970184326 & 0.7075069546699524 & \bfseries 0.7572644352912903 &  0.7270150184631348 & 0.7248818874359131 \\
		\midrule
		25.058237075805664 & 31.15333366394043 & 26.257211685180664 & \bfseries 31.433210372924805 & 31.355825424194336 \\
		\fpeval{100 * 0.07370038330554962} & \fpeval{100 * 0.015259397216141224} & \fpeval{100 * 0.049717504531145096} & \bfseries \fpeval{100 * 0.014507188461720943} & \fpeval{100 * 0.01571197621524334} \\
		0.6186882853507996 & 0.712673008441925 & 0.6861987709999084 & \bfseries 0.7252629399299622 & 0.7009592056274414 \\
		\midrule
		26.754220962524414 & 31.52347755432129 & 23.461977005004883 & 31.874441146850586 & \bfseries 32.09153366088867 \\
		\fpeval{100 * 0.05457334592938423} & \fpeval{100 * 0.014161386527121067} & \fpeval{100 * 0.09929478168487549} & \fpeval{100 * 0.01432089228183031} & \bfseries \fpeval{100 * 0.0124507499858737} \\
		0.7052920460700989 & 0.7458847165107727 & 0.6833181977272034 & \bfseries 0.7555358409881592 & 0.7461521029472351 \\
		\bottomrule
	\end{tabular}
	\hfill%
	\begin{tablenotes}\footnotesize
		\( \mathparagraph \): \gls{psnr} (\si{\decibel}) \( \uparrow \), \( \ddagger \): \gls{nmse} (\( \times \num{e2} \)) \( \downarrow \), \( \mathsection \): \gls{ssim} \( \uparrow \)
	\end{tablenotes}
\end{table*}
\begin{figure*}
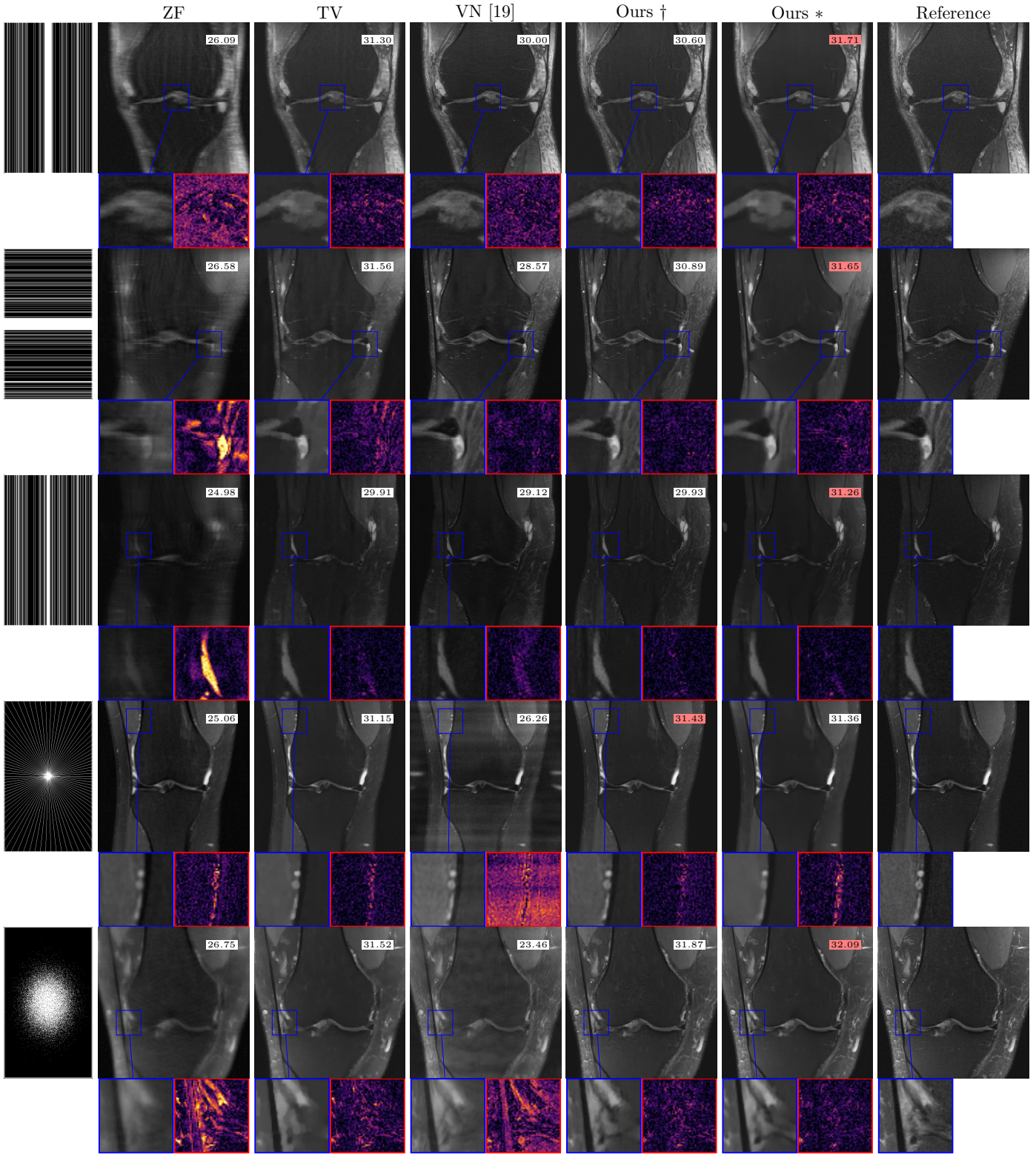

	\def\prefix{./figures/validation}
	\def\cherries{{"file1001916.h5","file1000528.h5","file1000344.h5","file1000000.h5","file1001643.h5"}}
	\def\spyoff{{{0.05,0.0},{0.7,-0.4},{-0.7,0.1},{-0.7,1.1},{-0.9,-0.4}}}
	\def\TableQuant{{
		{26.091176986694336, 31.302940368652344, 29.998945236206055, 30.602453231811523, 31.712961196899414 },
		{26.57789421081543, 31.557456970214844, 28.57035255432129, 30.890687942504883, 31.652633666992188},
		{24.9786319732666, 29.91111183166504, 29.117982864379883, 29.933029174804688, 31.255582809448242},
		{25.058237075805664, 31.15333366394043, 26.257211685180664, 31.433210372924805, 31.355825424194336},
		{26.754220962524414, 31.52347755432129, 23.461977005004883, 31.874441146850586, 32.09153366088867}
	}}
	\def\BestTag{{5, 5, 5, 4, 5}}
	\resizebox{\textwidth}{!}{%
	\begin{tikzpicture}
		\foreach [count=\isampling] \sampling in {cartesian, cartesian_rot90, {cartesian_4}, radial, gaussian2d}
		{
			\pgfmathsetmacro{\cherry}{\cherries[\isampling-1]}
			\foreach [count=\imethod] \method/\manno in {{zero_filled}/\gls{zf}, tv/TV, vn/{\gls{vn}~\cite{sriram_endtoend_2020}}, ours/{Ours \( \dagger \)}, ours/{Ours \( * \)},ground_truth/Reference}
			{
				\ifthenelse{\isampling=1}{
					\node at (3.1*\imethod, -2.8) {\manno};
				}{}
				\ifthenelse{\imethod=5}{
					\def\prefix{./figures/validation-fs-adapted}
				}{
					\def\prefix{./figures/validation-fs}
				}
				\pgfmathsetmacro{\spyxoff}{\spyoff[\isampling-1][0]}
				\pgfmathsetmacro{\spyyoff}{\spyoff[\isampling-1][1]}
				\coordinate (onn) at (\imethod * 3.1 + \spyxoff, -\isampling*4.5 + \spyyoff);
				\ifthenelse{\imethod=6}{}{
				\begin{scope}[spy using outlines={rectangle, magnification=3, width=1.47cm, height=1.47cm, connect spies}]
					\node at (3.1*\imethod, -\isampling*4.5) {\includegraphics[angle=180,origin=c,width=3cm]{\prefix/\sampling/\cherry/d_\method.png}};
					\spy [red] on (onn) in node [left] at (\imethod * 3.1 + 1.5 - 0.015, -\isampling*4.5-2.25);
				\end{scope}}
				\begin{scope}[spy using outlines={rectangle, magnification=3, width=1.47cm, height=1.47cm, connect spies}]
					\node at (3.1*\imethod, -\isampling*4.5) {\includegraphics[angle=180,origin=c,width=3cm]{\prefix/\sampling/\cherry/\method.png}};
					\spy [blue] on (onn) in node [left] at (\imethod * 3.1 - 0.015, -\isampling*4.5-2.25);
				\end{scope}
				\ifthenelse{\imethod<6}{
					\pgfmathsetmacro{\psnr}{\TableQuant[\isampling-1][\imethod-1]}
					\pgfmathsetmacro{\ibest}{\BestTag[\isampling-1]}
					\ifthenelse{\imethod=\ibest}{\def\ccolor{white!50!red}}{\def\ccolor{white}}
					\node [inner sep=0.1em, outer sep=0.1em, fill=\ccolor] at (3.1*\imethod+0.95, -\isampling*4.5+1.15) {\tiny\num[round-mode=places,round-precision=2]{\psnr}};
				}{}
			}
			\node at (0.6, -\isampling*4.5) {\includegraphics[cframe=white,height=3cm]{\prefix/\sampling/mask.png}};
		}
	\end{tikzpicture}}
	\caption{%
		Parallel imaging on out-of-distribution data (\( \dagger \): regularization parameters from non-fat-suppressed data, \( * \): adapted parameters):
		\nth{1} row:
		\num{4}-fold Cartesian sub-sampling with \qty{8}{\percent} \gls{acl},
		\nth{2} row:
		as previous with swapped phase encoding direction,
		\nth{3} row:
		\num{4}-fold Cartesian sub-sampling with \qty{4}{\percent} \gls{acl},
		\nth{4} row:
		Pseudo-radial sampling with \num{45} equidistant spokes (\( \approx \num{11} \)-fold sub-sampling).
		\nth{5} row:
		2D Gaussian sampling (\num{8}-fold sub-sampling).
		The inlays show the sub-sampling mask, a detail zoom (blue), and the magnitude of the difference to the reference (red, \num{0}~\protect\drawcolorbar~\num{0.2}).
	}%
	\label{fig:parallel imaging ood}
\end{figure*}

\begin{figure*}
	\resizebox{\textwidth}{!}{%
		\begin{tikzpicture}
			\foreach [count=\isens] \senslabel/\senspath in {%
				Ours/ours,%
				{ESPIRiT~\cite{uecker_espirit_13}}/espirit,%
				{\( \frac{\mathcal{F}^\ast(z_c)}{|(\mathcal{F}^\ast(z_c))_{c=1}^C|_\mathcal{C}} \)}/fully-sampled%
			} {%
				\node[rotate=90] at (-13.6,-\isens*3.2) {\senslabel};
				\node at (0, -\isens*3.2) {\foreach \idxx in {0,...,14}{%
					\includegraphics[rotate=180, width=.097\textwidth]{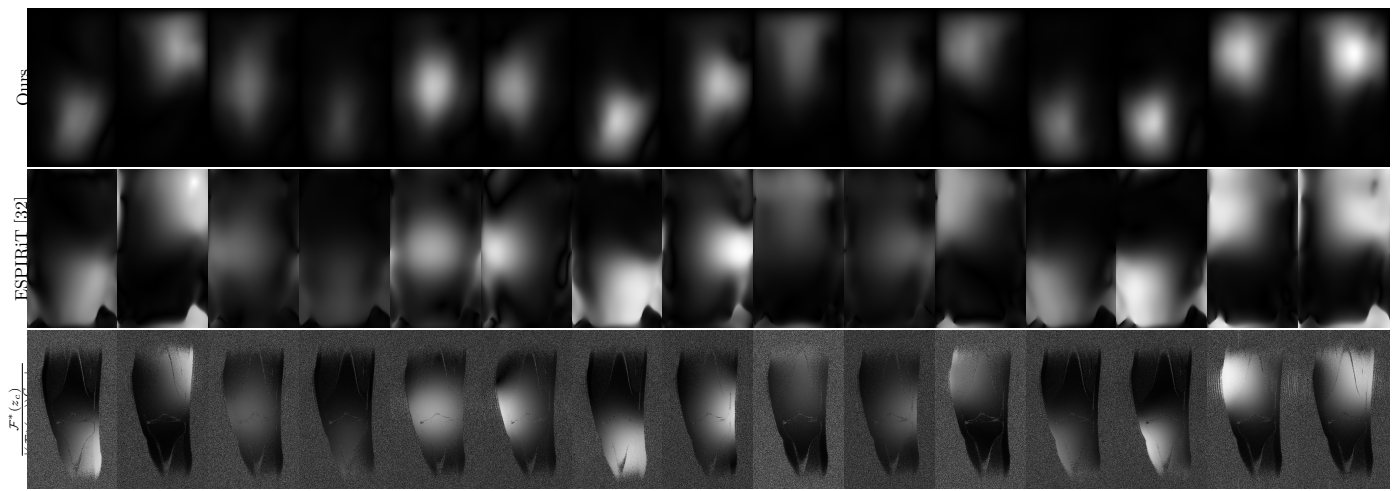}%
				}};
			}
		\end{tikzpicture}
	}%
	\caption{%
		\added{%
			Magnitude of the estimated sensitivities using our joint estimation algorithm (top) versus the ESPIRiT~\cite{uecker_espirit_13} estimation (middle) for the reconstruction problem shown in~\cref{fig:parallel imaging id} (\nth{1} row).
			The bottom row shows the reference coil sensitivities computed with the fully-sampled data.
		}
	}%
	\label{fig:sensitivities}
\end{figure*}

\added{%
	To analyze our reconstruction algorithm also with respect to the estimated sensitivity maps, we show example estimations along with ESPIRiT~\cite{uecker_espirit_13} in~\cref{fig:sensitivities}.
	The figure shows estimations for the \num{4}-fold Cartesian sub-sampling with \qty{8}{\percent} \gls{acl} reconstruction problem shown in the first row of~\cref{fig:parallel imaging id}.
	Due to the re-parametrization~\cref{eq:change of variables}, our reconstruction algorithm does not require pixel-wise normalized sensitivity maps.
	Hence, they look very physically plausible and match the reference well.
}

\begin{figure}
	\begin{tikzpicture}
		\node at (-2.1, 4.8) {\qty{8}{\percent} \gls{acl}};
		\node[rotate=90] at (-4.3, 2.3) {Ours};
		\node at (-2.1, 2.3) {\includegraphics[trim={0cm 4cm 0cm 4cm},clip,rotate=180,width=4cm]{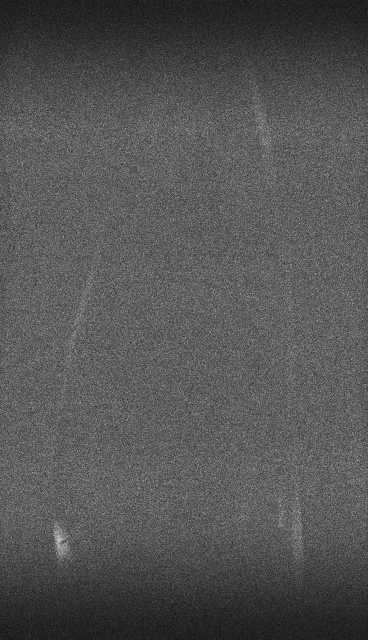}};
		\node at (2.1, 4.8) {\qty{4}{\percent} \gls{acl}};
		\node at (2.1, 2.3) {\includegraphics[trim={0cm 4cm 0cm 4cm},clip,rotate=180,width=4cm]{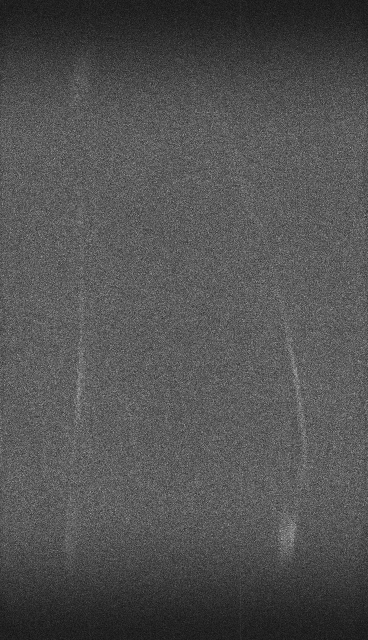}};
		\node[rotate=90] at (-4.3, -2.3) {ESPIRiT~\cite{uecker_espirit_13}};
		\node at (-2.1, -2.3) {\includegraphics[trim={0cm 4cm 0cm 4cm},clip,rotate=180,width=4cm]{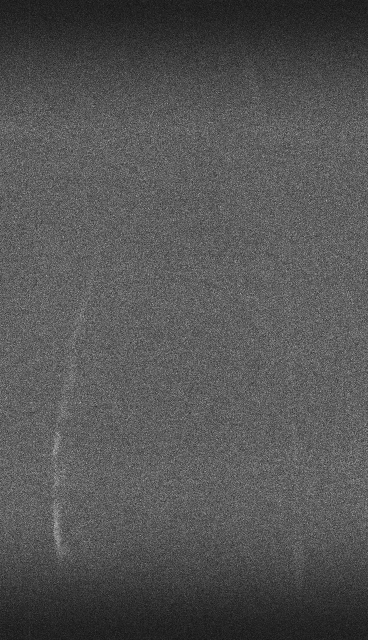}};
		\node at (2.1, -2.3) {\includegraphics[trim={0cm 4cm 0cm 4cm},clip,rotate=180,width=4cm]{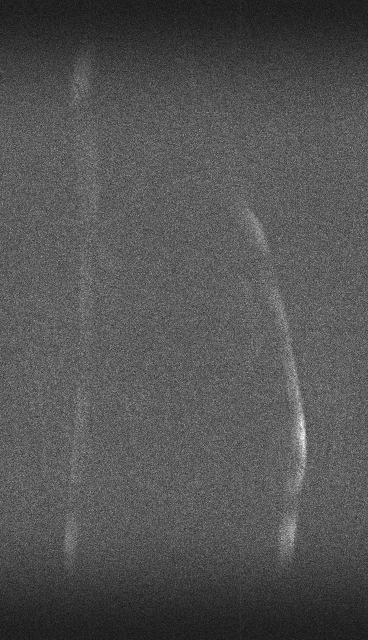}};

		\draw[-latex, line width=1mm] (-2.1, 1) -- (-2.9, .3);
		\draw[-latex, line width=1mm] (-2.1, -3.6) -- (-2.9, -4.3);

		\draw[-latex, line width=1mm] (1.8, 2.7) -- (1.0, 3.3);
		\draw[-latex, line width=1mm] (1.8, -1.9) -- (1.0, -1.3);
	\end{tikzpicture}
	\caption{%
			\gls{rss} null-space residual \(|(\pi_c)_{c=1}^C|_\mathcal{C} \) using our joint estimation algorithm (top) and ESPIRiT~\cite{uecker_image_2008} (bottom).
			We consider the two Cartesian sub-sampling reconstruction problems in the first and third row of~\cref{fig:parallel imaging id}: \qty{8}{\percent} \gls{acl} (left) and \qty{4}{\percent} \gls{acl} (right).%
	}%
	\label{fig:nullspace}
\end{figure}

\added{%
	We additionally analyze the estimation qualitatively by visualizing the \gls{rss} null-space residual~\cite{uecker_espirit_13} \( |(\pi_c)_{c=1}^C|_\mathcal{C} \).
	In~\cref{fig:nullspace}, we compare the sensitivity maps from our joint estimation with ESPIRiT for Cartesian sub-sampling patterns with \qty{8}{\percent} \gls{acl} and \qty{4}{\percent} \gls{acl} (first and third row in~\cref{fig:parallel imaging id} respectively).
	For both tasks, the sensitivity maps from our joint estimation algorithm can reproduce the data very well.
	In particular, while the ESPIRiT estimation leads to slightly better results when \qty{8}{\percent} \gls{acl} are available, the estimation deteriorates for \qty{4}{\percent}.
	In contrast, our estimation remains stable, as it can exploit data that is not in the \( k \)-space center.
}
\subsection{Generalization}
\label{ssec:generalization}
By construction, the learned model encodes the distribution of the training data.
In the previous sections we have demonstrated thoroughly that the prior is agnostic to shifts in the sub-sampling pattern, which is an expected consequence of the generative approach.
However, a natural question is whether the approach is also robust with respect to shifts in the underlying distribution.
To study this, we apply the regularizer to an underlying distribution of fat-suppressed images.

The quantitative and qualitative results in~\cref{tab:parallel imaging} and~\cref{fig:parallel imaging ood} indicate that our method generalizes better to unseen data \replaced{than}{that} the end-to-end \gls{vn} approach of~\cite{sriram_endtoend_2020}, although the performance degrades significantly in both cases.
To highlight the advantage of the tuneable regularization parameters in our approach, we show results using the \( \lambda \)-fit (see~\cref{ssec:methods pi}) calculated on non-fat-suppressed data, as well as using parameters adapted to the task:
The ability to tune the influence of the regularizer leads to improved performance when confronted with previously unseen data.
\section{Discussion}%
\label{sec:discussion}
\added{%
	\cref{fig:pdf sampling} demonstrates that our learned regularizer encodes the training distribution.
	As a consequence, we can reconstruct images from severely ill-posed problems (such as random sub-sampling pattern) satisfactorily.
	However, this also means that the performance of our learned regularizer is expected to significantly decline when applied to, e.g., different anatomy (rotation is explored in~\cite{zach_generative_2021}).
	Diffusion models~\cite{chung_scoremri_2022} are another instance of models that are capable of generating realistic-looking images from the data-distribution, but for inverse problems it is not clear if this is needed.
	Since our network is purely convolutional, it can easily be \enquote{localized} by removing the deeper layers, effectively recovering the fields-of-experts model~\cite{roth_fields_2005}.
	This would remove the majority of the trainable parameters (and consequently the training cost), and we hypothesize that it would lead to better generalization as local features are largely shared between different anatomical structures.
}

The results in~\cref{ssec:parallel imaging} demonstrate that the regularizer can reconstruct images \added{from multi-coil data} satisfactorily, irrespective of the acquisition mask.
This is a very significant advantage over the reference methods, since in practice a particular situation might necessitate a different \replaced{sub-sampling pattern}{acquisition protocol}.
As an example, phase encoding direction is often swapped when a blood vessel is located in a particularly disadvantageous position.
In addition, our regularizer does not need re-training when a new \added{sub-}sampling pattern is discovered to be advantageous.

\added{%
	The results on multi-coil data also empirically demonstrate that a real-valued regularizer suffices to reconstruct the underlying spin-density, at least for the acquisition sequences used in our dataset~\cite{zbontar_fastmri_2018}.
	Similarly to~\cite{song2022solving,luo_bayesian_2023}, our approach could easily be extended to account for complex images if needed.
	In particular, this is the case for contrasts that inherently rely on complex-valued information, such as phase-contrast \gls{mri}.
}

Generalization is not limited to the acquisition mask:
The regularizer also performs well in out-of-distribution experiments, which highlight the importance of the ability to control its influence.
\cref{tab:parallel imaging} shows that adapting the strength of the regularization to the underlying data strongly improves performance.
We note that the regularization strength was always tuned to a \num{4}-fold Cartesian sub-sampling task with \qty{8}{\percent} \gls{acl} (see~\cref{ssec:methods pi}).
The results for radial sub-sampling indicate that this is a sub-optimal fit for the out-of-distribution data:
The fit calculated on the \gls{corpd} data (\( \dagger \) column) performs better than the fit calculated on \gls{corpdfs} data.
\cref{fig:parallel imaging ood} \deleted{(right)} suggests that the adapted parameters lead to an over-smoothed reconstruction.
Clearly, the performance could be improved by adapting the regularization strength to the \gls{corpdfs} data \emph{and} the radial sub-sampling.
In general, adapting the regularization strength to acquisition mask as well as the data is greatly beneficial.
However, this is typically not possible with other data-driven approaches.

The accompanying probabilistic interpretation is significant in two ways:
First, the regularizer can be inspected by data-independent analysis as in~\cref{fig:pdf sampling}, where experts can easily visualize preferred structures.
This aids in interpreting the encoded information and thus improves the confidence with which clinicians may view the reconstruction.
Second, the variance maps provide the clinician with additional information about the inherent uncertainty in the reconstruction.
\cref{fig:simulation study} (right) clearly shows high variance around small anatomic structures.
If the clinician's decision making is based on such regions, they might deem it necessary to acquire additional information.

In the single-coil simulation study, quantitative and qualitative analysis shows superiority of the \gls{mmse} estimate over the \gls{map} estimate.
We believe that this is related to the training procedure:
During training, the regularizer is only ever confronted with slightly noisy images in the Langevin process (see~\cref{ssec:methods ml}), and injecting the same noise in the reconstruction algorithm improves overall performance.
We are unsure why this performance improvement does not translate to the parallel imaging experiments, but hypothesize that the our joint estimation biases the reconstruction such that this effect is no longer observable.

Finally, we emphasize that our algorithm for reconstructing parallel \gls{mri} is fast:
\cref{alg:ipalm} converges in around \num{100} iterations, which takes around \qty{5}{\second} on an NVIDIA Titan RTX using approximately \qty{2}{\giga\byte} of memory.
This is in stark contrast to score-based diffusion models, where the reverse diffusion process typically has to be solved with high accuracy and consequently reconstruction time is in the order of \qty{10}{\minute}~\cite{chung_scoremri_2022}.
Speed is advantageous in practice, as images can be viewed while the patient is still in the scanner.
This allows fast changes to the \replaced{sub-sampling pattern}{acquisition protocol}, should they be needed.
\section{Conclusion}
We utilize modern \replaced{generative}{unsupervised} learning techniques to train a regularizer such that it encodes the underlying distribution of the training data faithfully.
By embedding the regularizer in a variational reconstruction framework, we can satisfactorily reconstruct single-coil and parallel \gls{mri} by adapting the data likelihood term.
Quantitative and qualitative analysis indicate competitive reconstruction performance, on-par with (or sometimes superior to) fully supervised methods.
In addition, our approach is agnostic to changes in the acquisition mask, while the supervised reference methods introduce severe hallucinations when confronted with previously unseen data.
The knowledge of the human anatomy encoded in the regularizer even allows us to reconstruct high quality images from random sampling masks.
Finally, on out-of-distribution data, our method is still on-par or better than hand-crafted regularizers such as the \gls{tv} or other supervised methods.
This indicates that our method also generalizes better with respect to shifts in the underlying distribution.

In addition to a competitive reconstruction performance, our method has a natural accompanying probabilistic interpretation through statistical modeling and Bayesian inference.
This allows experts to explore a distribution of reconstructions, while other data-driven approaches only act as point estimators.
Our experiments indicate that this distribution encodes important diagnostic information, such as high uncertainty around small anatomical structures.
We believe that this additional information can aid the practitioner's decision making.
Further, we can perform a data-independent analysis of the model by visualizing the induced distribution.
This allows to interpret the information encoded in the regularizer, whereas other data-driven methods solely act as black-boxes in this respect.

To reconstruct parallel \gls{mri}, we propose a fast algorithm to jointly estimate the image and sensitivity maps.
In contrast to off-line sensitivity estimation, our approach does not require auto-calibration lines, can easily be applied to non-Cartesian sampling trajectories, and uses all available data.
The resulting optimization problem can be solved efficiently using non-convex optimization algorithms.
High-fidelity reconstructions are available in approximately \qty{5}{\second} on consumer hardware.

In general, we believe that reconstruction approaches based on generative priors have huge potential.
A natural extension of our work would be to combine the image-prior with a learned sensitivity-prior.
In addition, future research could focus on extending the simple architecture used in this paper with more modern building blocks, such as attention layers.
\added{%
	A related direction is to research if local convolutional models can replicate the performance of our generative prior.
}
\appendix
\section*{Appendix}
\subsection*{Spline fit}
\added{
\cref{fig:splines} shows the scatter plot of reconstructed versus reference intensities along with the spline fit discussed in~\cref{ssec:parallel imaging}.
\begin{figure}
	\includegraphics[trim={0cm 0cm 2cm 0cm}, clip, width=\columnwidth]{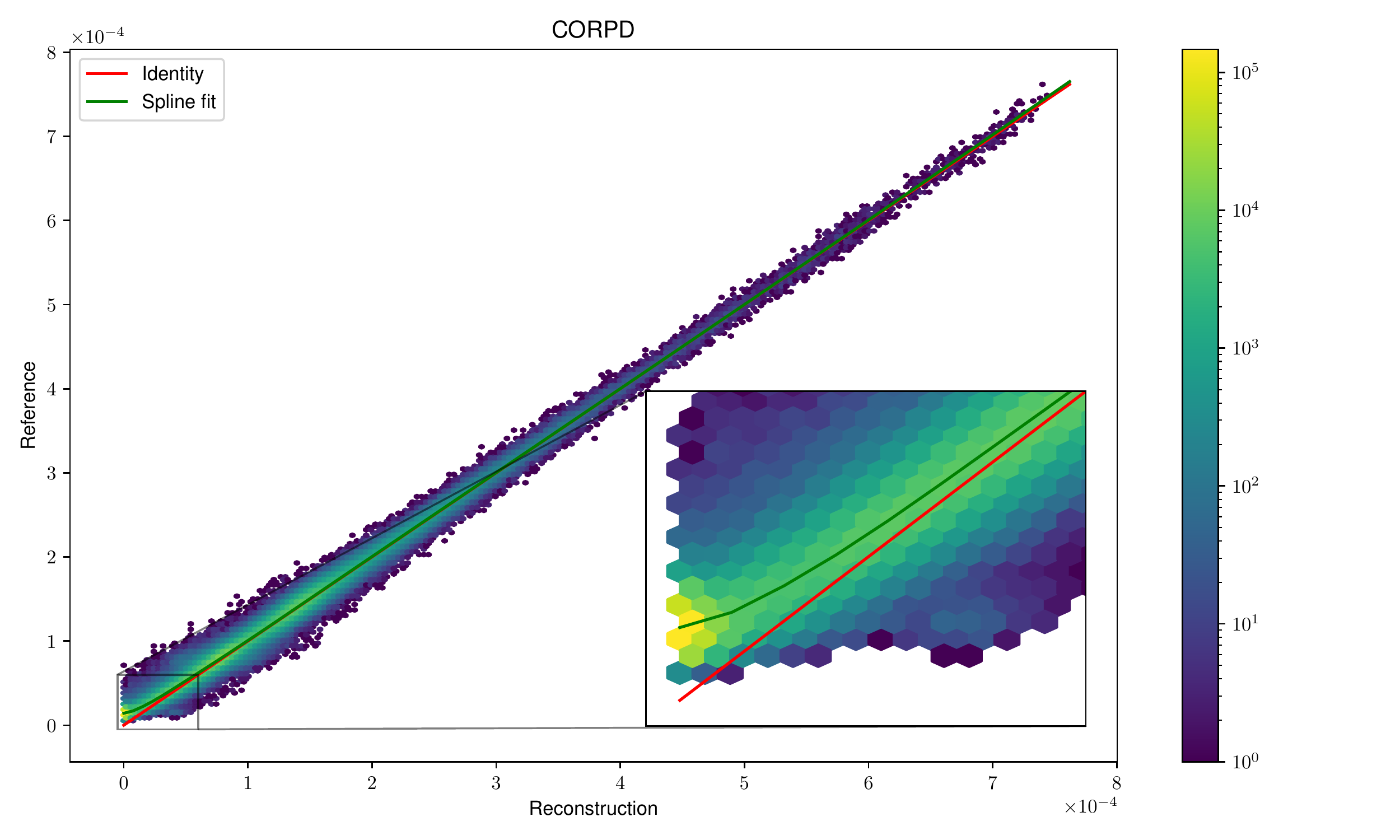}
	\includegraphics[trim={0cm 0cm 2cm 0cm}, clip, width=\columnwidth]{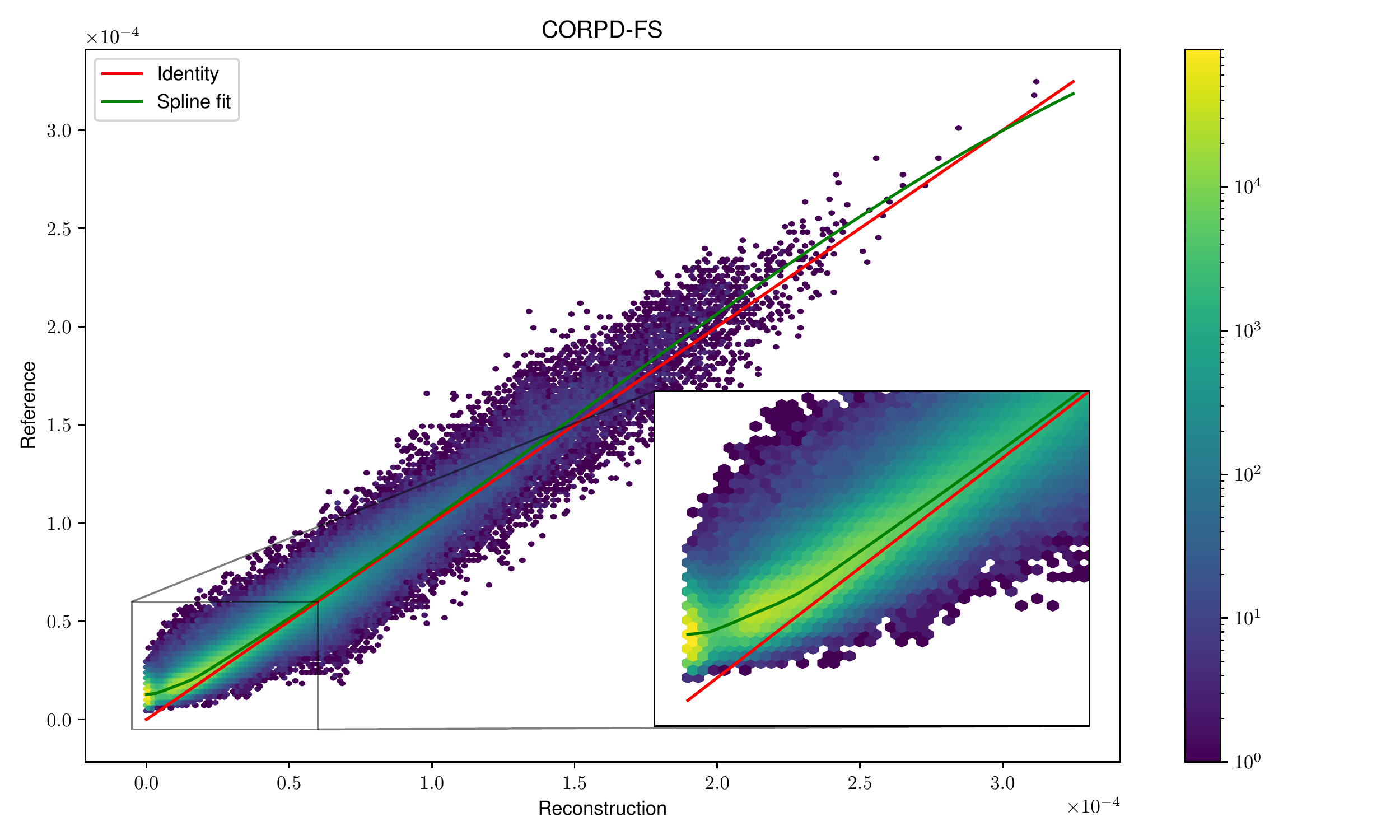}
	\caption{%
		\added{Spline fit computed on a validation set for CORPD (top) and CORPD-FS data (bottom).
		The figures show the log-histogram of reconstructed versus reference intensities with the insets show the region around zero, where they deviate the strongest.}
	}
	\label{fig:splines}
\end{figure}
}
\subsection*{Additional results}
\begin{figure}
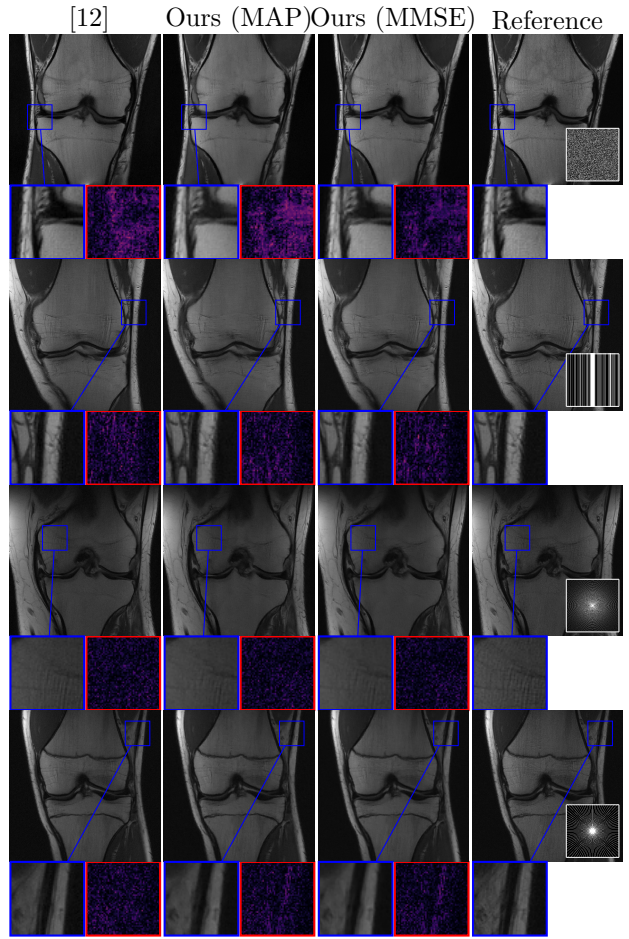

	\centering
	\def\spyoff{{%
		{-0.6,-0.1},%
		{0.65,0.3},%
		{-0.4,0.3},%
		{.7,.7}}%
	}
	\begin{tikzpicture}
		\def\prefix{./figures/single-coil-synthetic}
		\foreach [count=\ipattern] \subpattern in {random, cartesian, spiral, radial}
		{
			\pgfmathsetmacro{\spyxoff}{\spyoff[\ipattern-1][0]}
			\pgfmathsetmacro{\spyyoff}{\spyoff[\ipattern-1][1]}
			\foreach [count=\iwhich] \which/\anno in {recon/~\cite{chung_scoremri_2022}, ours/Ours (\gls{map}), mean/Ours (\gls{mmse}), ground_truth/Reference}
			{
				\ifthenelse{\ipattern=1}{\node at (2.05*\iwhich, -1.8) {\anno};}{}
				\coordinate (onn) at (\iwhich * 2.05 + \spyxoff, -\ipattern*3 + \spyyoff);
				\ifthenelse{\iwhich=4}{}{
				\begin{scope}[spy using outlines={rectangle, magnification=3, width=0.98cm, height=0.98cm, connect spies}]
					\node at (2.05*\iwhich, -\ipattern*3) {\includegraphics[angle=180,origin=c,width=2cm]{./figures/diffusion-comparison/\subpattern/d_\which.png}};
					\spy [red] on (onn) in node [left] at (\iwhich * 2.05 + 1, -\ipattern*3-1.5);
				\end{scope}}
				\begin{scope}[spy using outlines={rectangle, magnification=3, width=0.98cm, height=0.98cm, connect spies}]
					\node at (2.05*\iwhich, -\ipattern*3) {\includegraphics[angle=180,origin=c,width=2cm]{./figures/diffusion-comparison/\subpattern/\which.png}};
					\spy [blue] on (onn) in node [left] at (\iwhich * 2.05-0.01, -\ipattern*3-1.5);
				\end{scope}
			}
			\node at (2.65 + 2.05*3, -\ipattern*3-.6) {\includegraphics[cframe=white,height=.7cm]{\prefix/\subpattern/mask.png}};
		}
	\end{tikzpicture}
	\caption{%
		\added{%
			Qualitative comparison against~\cite{chung_scoremri_2022} on in-distribution data:
			\nth{1} row:
			Random sub-sampling with acceleration factor \num{3},
			\nth{2} row:
			\num{4}-fold Cartesian sub-sampling with \qty{8}{\percent} \gls{acl},
			\nth{3} row:
			Spiral sub-sampling (\num{5}-fold acceleration),
			\nth{4} row:
			Radial sub-sampling using \num{45} spokes.
			The inlays show the sub-sampling pattern (white), a detail zoom (blue), and the magnitude of the difference to the reference (red, \num{0}~\protect\drawcolorbar~\num{0.2}).
		}
	}%
	\label{fig:diffusion comparison}
\end{figure}

\added{%
	A qualitative comparison against the diffusion-based method of~\cite{chung_scoremri_2022} is presented in~\cref{fig:diffusion comparison}.%
	The accompanying quantitative evaluation is shown in~\cref{tab:diffusion comparison} in the main body.
}
\section*{Acknowledgment}
We thank the reviewers for their time and effort spent on reviewing the manuscript.
Their comments and suggestions helped us to greatly improve the quality the manuscript.
\printbibliography%
\end{document}